\documentclass[a4paper,unpublished,onecolumn,10pt]{quantumarticle}
\pdfoutput=1

\usepackage[numbers,sort&compress]{natbib}
\usepackage{amsfonts}
\usepackage{amssymb}
\usepackage{amsmath}
\usepackage{amsthm}
\usepackage{graphicx}
\usepackage{verbatim}
\usepackage{hyperref}
\usepackage[english]{babel}
\usepackage[pdftex]{color}
\usepackage[dvipsnames]{xcolor}

\newtheorem{theorem}{Theorem}
\newtheorem{proposition}{Proposition}
\newtheorem{definition}{Definition}

\newtheorem{corollary}{Corollary}

\hypersetup{
    colorlinks=true,       % false: boxed links; true: colored links
    linkcolor=cyan,          % color of internal links
    citecolor=magenta,        % color of links to bibliography
    filecolor=magenta,      % color of file links
    urlcolor=cyan,           % color of external links
    runcolor=cyan
}

\graphicspath{{figs/}}

\newcommand{\bra}[1]{\langle#1|}
\newcommand{\ket}[1]{|#1\rangle}
\newcommand{\braket}[2]{\langle#1|#2\rangle}
\newcommand{\ketbra}[2]{|#1\rangle \langle #2 |}

\begin{document}

\title{Simulation of quantum optics by coherent state decomposition}

\author{Jeffrey Marshall}\thanks{jmarshall@usra.edu}
\affiliation{QuAIL, NASA Ames Research Center, Moffett Field, CA 94035, USA}
\affiliation{USRA Research Institute for Advanced Computer Science, Mountain View, CA 94043, USA}

\author{Namit Anand}\thanks{namit.anand@us.kbr.com}
\affiliation{QuAIL, NASA Ames Research Center, Moffett Field, CA 94035, USA}
\affiliation{KBR, Inc., 601 Jefferson St., Houston, TX 77002, USA}

\begin{abstract}
We introduce a framework for simulating quantum optics by decomposing the system into a finite rank (number of terms) superposition of coherent states.
This allows us to define a resource theory, where linear optical operations are `free' (i.e., do not increase the rank), and the simulation complexity for an $m$-mode system scales quadratically in $m$, in stark contrast to the Hilbert space dimension.
We outline this approach explicitly in the Fock basis, relevant in particular for Boson sampling, where the simulation time (space) complexity for computing output amplitudes, to arbitrary accuracy, scales as $O(m^2 2^n)$ ($O(m2^n)$), for $n$ photons distributed amongst $m$ modes. 
We additionally demonstrate that linear optical simulations with the $n$ photons initially in the same mode scales efficiently, as $O(m^2 n)$.
This paradigm provides a practical notion of `non-classicality', i.e., the classical resources required for simulation. Moreover, by making connections to the stellar rank formalism, we show this comes from two independent contributions, the number of single-photon additions, and the amount of squeezing.
\end{abstract}

\maketitle

\tableofcontents

\section{Introduction}
The classical simulation of quantum mechanics is a critically important field for science in general. Not only does this pursuit allow us to push the boundaries of what is possible to simulate, it also enables new insights into notions of complexity and information theory. The field of quantum computing itself has its origins in classical simulation, with early observations by Feynman and Manin that quantum mechanics, seemingly, is incompatible with classical mechanics, in a particular computational sense. Indeed, the advent of quantum complexity theory and several pioneering works in the 1990's suggested that the class of problems efficiently computable on a (randomized) classical computer (BPP) is a strict subset of those efficiently computable on a quantum computer (BQP), up to standard complexity theoretic assumptions \cite{aaronson2009bqp}.

Quantum simulation has itself been driven by experimental developments in the field of quantum computing, for as sizes in the lab increase, better and better classical simulation tools are required for verification \cite{arute_quantum_2019, mi_information_2021, morvan2023phase}. As a result, a plethora of techniques have been developed for such purposes in finite dimensional systems, including state vector simulation \cite{quantum_ai_team_and_collaborators_2020_4023103}, stabilizer state simulation \cite{aaronson-gottesman-stabilizer, bravyiImprovedClassicalSimulation2016, Bravyi2019simulationofquantum}, tensor network methods \cite{vidal-efficient-2003, markov-tensor-contract, markov_quantum_2018, pan_contracting_2020}, and many other optimized routines within these broad classes \cite{pednault2019leveraging, villalonga_flexible_2019, gray_hyper-optimized_2021, mandra_hybridq_2021}.

Infinite dimensional systems are typically treated quite distinctly from finite dimensional ones in terms of simulation, although there are certain similarities. For example, a generalization of the Clifford group to infinite dimensions, via the Weyl relations, allows for a stabilizer formalism to be employed \cite{efficient-cv}, which shows that Gaussian computations can be simulated efficiently, in a similar sense to the Gottesman-Knill theorem in finite dimensions \cite{positive_wigner_sim-Eisert, Veitch_2013}. 
Conversely, the phase space formalism, originally developed for continuous variable (CV) systems \cite{wigner}, can also be defined in finite dimensions, which connects the negativity of the discrete Wigner function to the ability to attain a `quantum speedup' \cite{veitch_negative_2012, wigner-rebits, raussendorf_phase-space-simulation_2020}.
Additionally, and more recently, tensor network methods for Bosonic systems have also been explored \cite{liu2023complexity, cilluffo_simulating_2023, oh_tensor_2023}. 
Improvements to non-Gaussian simulation continue to increase the threshold for a quantum advantage \cite{ 20-photon-fock-BS, GBS-2021, boson-sampling-advantange}, and our understanding of the intrinsic complexity of simulating these systems \cite{aaronson-boson-sampling, clifford-clifford-2017, clifford-clifford-2020}. We review other relevant simulation techniques in Sect.~\ref{sect:background}.

\textit{\textbf{Contributions}}.--- In this work we outline an approach to the simulation of infinite dimensional systems, inspired by stabilizer rank simulations in finite dimensional systems with `magic' \cite{Bravyi2019simulationofquantum}. 
First we show (Th.~\ref{th:general-fock}) that any state in the Fock basis can be decomposed, to arbitrary accuracy, as a finite superposition of coherent states.
Then, with such a decomposition we demonstrate that the cost of classically simulating linear optics (Th.~\ref{th:lo-free}), and Fock basis measurements (Sect.~\ref{sect:measurements}), scale linearly with respect to the number of terms in the decomposition.
This results in a simulation time complexity for computing Fock state amplitudes in the setting of Boson sampling, scaling as $O(m^2 2^n)$ for $n$ photons in $m$ modes, scaling similarly in $n$ with techniques relying upon permanent calculations (though the memory overhead in our approach is worse, $O(m2^n)$ vs. $O(m)$) \cite{aaronson-boson-sampling, clifford-clifford-2017}. 
Additionally we note that this approach can simulate bunched $n$ photon systems $\ket{n}\ket{0}^{\otimes m-1}$ in a time that scales linearly in the number of photons, $O(nm^2)$ (note, permanent computations also reduce to similar complexity in this case). This allows up to a logarithmic number of N00N states $(\ket{N0} + \ket{0N})$ -- fundamental for quantum metrology and sensing \cite{dowling-noon} -- to be simulated efficiently, in a linear optical setting.
Moreover, the framework is flexible in the sense it can, in principle, simulate both linear and non-linear optics, finite photon number states, coherent states, and squeezed states in the same manner (though these do not all have the same simulation complexity, i.e., some are less efficient).

In Sect.~\ref{sect:discussion} we also discuss trade offs that can be employed, interchanging accuracy (fidelity) for simulation cost (memory and time). Further, we outline an approach that can be applied to circuits with structure, partitioning the system according to the number of entangling operations, resulting in a time and memory saving for certain simulations.

This work provides a well defined notion of classical complexity for the simulation of infinite dimensional quantum systems, related to the efficiency of a decomposition to coherent states. 
Indeed, coherent states are often considered the `most classical' of quantum states \cite{gerry_knight_2004}, and the simulation cost in this framework is linear in the number of these classical states.
As such we construct a quantum resource theory for coherent state basis simulations (Sect.~\ref{sect:resource_th}), which further relates this `non-classicality' to i) the number of photon additions and ii) the amount of squeezing required to construct a state. We compare this resource to the non-Gaussianity measures, the Wigner negativity, and the stellar rank.

\section{Background \label{sect:background}}

\subsection{Hilbert space, operators and states}
Here we introduce necessary prerequisites and notation for what is to follow in Sect.~\ref{sect:sim}. The starting point is a separable infinite dimensional Hilbert space, $\mathcal{H}_\infty$, such as the Hilbert space for a particle in one dimension (the space of square integrable functions over the real line). By definition, such a Hilbert space admits a countable orthonormal basis, $\mathcal{H}_\infty=\mathrm{span}\{\ket{n}\}_{n=0}^\infty$, which we will call the Fock basis. Each basis element, $\ket{n}$, will be called a single mode Fock state (or simply Fock state). Although we have not specified the system per se, for ease of language, we will refer to $\ket{n}$ as the Fock state of $n$ photons. Also note, that whilst in this section we will focus on the single mode case, the discussion generalizes straightforwardly to the multi-mode case, $\mathcal{H} = \mathcal{H}_\infty^{\otimes m} = \mathrm{span}\{\ket{n_1, \dots, n_m}\}_{n_i=0}^\infty$, where $m$ is the number of modes.

With the basis in hand, we can construct relevant operators for quantum optics. In particular we define the annihilation and creation operators,
\begin{equation*}
    \hat{a} = \sum_{n=1}^\infty \sqrt{n} \ketbra{n-1}{n},\,\hat{a}^\dag = \sum_{n=0}^\infty \sqrt{n+1}\ketbra{n+1}{n}.
\end{equation*}
Whilst these operators themselves are not Hermitian, they constitute the building blocks for all observable operators of interest in the present work.
We will define these concisely below, and then discuss some of their properties:
\begin{enumerate}
    \item The number operator: $\hat{n}=\hat{a}^\dag \hat{a} = \sum_{n=0}^\infty n\ketbra{n}{n}$
    \item The `position' operator: $\hat{q} = (\hat{a}^\dag + \hat{a})/2$
    \item The `momentum' operator: $\hat{p} = i(\hat{a}^\dag - \hat{a})/2$
    \item The displacement operator\footnote{The second relation follows from the BCH formula, noting $[\hat{a}, \hat{a}^\dag]=1$ \cite{fisher_impossibility_1984}.}: $\hat{D}(\alpha) = e^{\alpha \hat{a}^\dag - \bar{\alpha}\hat{a}}=e^{-|\alpha|^2/2}e^{\alpha \hat{a}^\dag}e^{-\bar{\alpha}\hat{a}}$, where $\alpha \in \mathbb{C}$, and $\bar{z}$ is the complex conjugate of $z$
\end{enumerate}

In this work, we will refer to $\hat{q}, \hat{p}$ as the position and momentum operators (or `quadratures'), however there may be no connection to such physical notions, as we've not fixed ourselves to any particular system (for example, in the theory of superconductivity these are the flux and charge operators respectively \cite{qed-review}).
Indeed, the framework we are presenting is quite general, and applies to \textit{any} infinite dimensional quantum system. The choice of basis is also not particularly important, as we can always perform a unitary rotation on the basis $\ket{n} \rightarrow \hat{U}\ket{n}$ and define our operators with respect to this, $\hat{a} \rightarrow \hat{U}\hat{a}\hat{U}^\dag$, etc. (though in practice, often there will be a `canonical' choice, coming from the physics).
It is also worth noting, some conventions differ in $\hat{q}, \hat{p}$, with common choices replacing the denominator (2) by $\sqrt{2}$ or $1$.

These operators have other remarkable properties. Namely, it is easy to show they admit a non-trivial commutation relation $[\hat{q}, \hat{p}]=i/2$ (where the right hand side has an implicit identity operator)\footnote{Notice the commutation relation $[\hat{q}, \hat{p}]=i/2$ immediately implies $\hat{q}, \hat{p}$ are not trace-class.}. It is further possible to show these operators have a continuous and real spectra \cite{GaussianQI-Weedbrook}, i.e.,
\begin{equation*}
    \hat{q}\ket{q} = q \ket{q};\;\;\hat{p}\ket{p} = p \ket{p},
\end{equation*}
where $q,p\in \mathbb{R}$. It is important to observe that, whilst $\ket{q}, \ket{p}$ are often treated as idealized quantum states (and approximations can be made that are arbitrarily close), they are in fact not legitimate quantum states; in particular, the wavefunction is a delta function and not square integrable, $\langle q|q'\rangle = \delta(q-q')$. 

This brings our attention to the displacement operator, $\hat{D}(\alpha)$. This can be used to construct an uncountably infinite family of quantum states, known as coherent states. In particular, by starting at the so called vacuum state $\ket{0}$ (the Fock state of 0 photons) and applying the displacement operator, one arrives at the coherent state
\begin{equation}
    \ket{\alpha} := \hat{D}(\alpha)\ket{0} = e^{-|\alpha|^2/2}e^{\alpha \hat{a}^\dag}\ket{0} = e^{-|\alpha|^2/2}\sum_{n=0}^\infty \frac{\alpha^n}{\sqrt{n!}}\ket{n}.
    \label{eq:coherent-state}
\end{equation}
Note, the Fock state $\ket{0}$ is also a coherent state (i.e. the one with $\alpha=0$), though this is not true for any other Fock state $n>0$ (it should be clear from the context and notation (e.g., Latin vs. Greek symbols) when we are referring to a Fock or coherent state in this work). It is easy to check Eq.~\eqref{eq:coherent-state} defines a genuine normalized quantum state, with expected photon number $\langle \alpha|\hat{n}|\alpha\rangle = |\alpha|^2$.

Coherent states themselves have several useful and interesting properties. First, they form a basis over the Hilbert space
\begin{equation}
    \frac{1}{\pi}\int d\alpha \ketbra{\alpha}{\alpha} = \mathbb{I},
    \label{eq:coherent-resolution-identity}
\end{equation}
and as such any state can be represented 
\begin{equation}
    \ket{\psi} = \frac{1}{\pi} \int d\alpha \psi(\bar{\alpha}) \ket{\alpha},
    \label{eq:coherent-integral}
\end{equation}
where the integral is evaluated over the plane, $d\alpha = d\mathrm{Re}(\alpha)d\mathrm{Im}(\alpha)$, and $\psi(\bar{\alpha})=\braket{\alpha}{\psi}$ \cite{goldbergExtremalQuantumStates2020}. 
Unlike Fock states, coherent states form an uncountable, overcomplete basis for the Hilbert space \(\mathcal{H}_{\infty}\) and have the following inner product formula,
\begin{equation}
    \langle \alpha |\beta\rangle = e^{-\frac{1}{2}|\alpha - \beta|^2}e^{-i \mathrm{Im}(\alpha \bar{\beta})}.
    \label{eq:coherent-overlap}
\end{equation}
Coherent states are also the eigenstates of the annihilation operator: $\hat{a}\ket{\alpha} = \alpha \ket{\alpha}$.
Experimentally, the value $\alpha$ of a coherent state can be determined by `homodyne' measurement, i.e., measuring the position/momentum operators: $\langle \alpha|\hat{q}|\alpha\rangle = \mathrm{Re}(\alpha)$, $\langle \alpha|\hat{p}|\alpha\rangle = \mathrm{Im}(\alpha)$. 

Coherent states are often referred to as the `most classical' states of a CV system \cite{gerry_knight_2004}. This is because they minimize the uncertainty relations for both quadratures, \(\hat{q},\hat{p}\), with equal variance in either observable. Squeezed states on the other hand also minimize the uncertainty relation, while \textit{not} equalizing the variance of each observable. For example, one could minimize the position uncertainty at the cost of increasing the momentum uncertainty. Define the ``squeeze'' operator as \cite{fisher_impossibility_1984}
\begin{align}
\begin{split}
& \hat{S}(\zeta):= \exp \left[ \frac{1}{2} \left( \overline{\zeta} \hat{a}^{2} - \zeta \hat{a}^{\dagger 2}  \right) \right] = \\ & \frac{1}{\sqrt{\cosh r}} \exp\left[-\frac{1}{2}e^{i\phi} \tanh(r) \hat{a}^{\dag 2}\right]\exp\left[-\ln(\cosh(r))\hat{n}\right]\exp\left[\frac{1}{2}e^{-i\phi}\tanh(r) \hat{a}^2\right],
\end{split}
\label{eq:squeeze-operator}
\end{align}
where \(\zeta = r e^{i \phi}\) is a complex number known as the squeeze parameter, with \(0 \leq r < \infty\) and \(0 \leq \theta < 2 \pi\). We will refer to \(| \zeta \rangle := \hat{S}(\zeta) | 0 \rangle\) as squeezed vacuum states, although one can also consider squeezing other coherent states instead of the vacuum (we will reserve the symbol $\zeta$ to denote squeezed states). Any arbitrary pure single-mode Gaussian state \(| \psi \rangle\) can be obtained by displacing a squeezed vacuum state, or squeezing a coherent state, i.e., \(| \psi \rangle = \hat{D}(\alpha)\hat{S}(\zeta)\ket{0} = \hat{S}(\zeta) \hat{D}(\gamma) | 0 \rangle\), where $\gamma = \alpha \cosh r + \bar{\alpha}e^{i\phi} \sinh r$.

Before delving into more details, it is worth listing some useful identities involving displacement and squeeze operators, and their canonical transformations. The displacement operators are unitary and form a one-parameter family,
\begin{align}
\label{eq:displacement-family}
\begin{split}
&\hat{D}(\alpha) \hat{D}^\dag (\alpha) = \mathbb{I} =  \hat{D}^\dag (\alpha) \hat{D}(\alpha),\\
& \hat{D}(\alpha) \hat{D}(\beta) = e^{i \mathrm{Im}(\alpha \overline{\beta})} \hat{D}(\alpha+\beta).
\end{split}
\end{align}
Moreover, they displace the annihilation and creation operators as,
\begin{align}
\label{eq:displacement-creation-annihilation}
\begin{split}
&\hat{D}^\dag (\alpha) \hat{a} \hat{D}(\alpha) = \hat{a} + \alpha = \hat{D} (-\alpha) \hat{a} \hat{D}^\dag(-\alpha),\\
&\hat{D}^\dag (\alpha) \hat{a}^{\dagger} \hat{D}(\alpha) = \hat{a}^{\dagger} + \overline{\alpha} = \hat{D} (-\alpha) \hat{a}^{\dagger} \hat{D}^\dag(-\alpha).
\end{split}
\end{align}
By unitarity, this generalizes to, $\hat{D}^\dag(\alpha)\hat{a}^n \hat{D}(\alpha) = (\hat{a} + \alpha)^n$, and $\hat{D}^\dag(\alpha)\hat{a}^{\dag n} \hat{D}(\alpha) = (\hat{a}^\dag + \bar{\alpha})^n$.

Similarly, the squeeze operator is unitary,
\begin{align}
\hat{S}(\zeta) \hat{S}^\dag (\zeta) = \mathbb{I} = \hat{S}^{\dagger}(\zeta) \hat{S}(\zeta),
\end{align}
and they transform the annihilation and creation operators as (hyperbolic rotations),
\begin{align}
\label{eq:squeeze-annihilation-creation}
\begin{split}
& \hat{S}^{\dagger}(\zeta) \hat{a} \hat{S}(\zeta) = \hat{a} \cosh(r) - \hat{a}^{\dagger} e^{i \phi} \sinh(r),\\
& \hat{S}^{\dagger}(\zeta) \hat{a}^{\dagger} \hat{S}(\zeta) = \hat{a}^{\dagger} \cosh(r) - \hat{a} e^{-i \phi} \sinh(r).
\end{split}
\end{align}
The squeeze operator does not commute with the displacement, but rather has an interesting (braiding type) relation
\begin{align}
\label{eq:squeeze-displacement-commutation}
\hat{D}(\alpha) \hat{S}(\zeta) = \hat{S}(\zeta) \hat{D}(\gamma),
\end{align}
where $\gamma = \alpha \cosh r + \bar{\alpha} e^{i\phi} \sinh r$ \cite{fisher_impossibility_1984}.

\subsection{QPD simulation \label{sect:efficient-sim}}
Central to the discussion of optics simulation is the Wigner function, which provides a phase space formalism for infinite dimensional quantum systems. Every state $\rho$ of $m$ modes has an associated Wigner function $W:\mathbb{C}^{m}\rightarrow \mathbb{R}$, which is informationally equivalent to the state. For a single mode state $\rho$ the Wigner function is defined as\footnote{The integral is over the entire plane $d\beta = d\mathrm{Re}(\beta)d\mathrm{Im}(\beta)$. Depending on the convention relating the creation/annihilation operators to the position/momentum operators, the Wigner function may have different normalization. Under our convention, for pure states, one can show $|W| \le 2/\pi$. }:
\begin{equation}
    W(\alpha) = \frac{1}{\pi^2}\int d\beta e^{\bar{\beta}\alpha - \beta \bar{\alpha}}\mathrm{Tr}[\hat{D}(\beta)\rho].
    \label{eq:wigner}
\end{equation}
This can be generalized to multi-mode states easily, see e.g. Ref.~\cite{GaussianQI-Weedbrook} and App.~\ref{sect:derviation-off-diagonal-wigner} for more information.
 Through this phase space formalism, evolution dynamics, expectation values and measurement outcomes can be computed from the Wigner function. The Wigner function of a state defines a quasi-probability distribution (QPD), since it is normalized in the sense $\int d\alpha W(\alpha) = \mathrm{Tr}[\rho]$, but $W$ can attain negative values in general.

The Wigner function for a coherent state $\ket{\beta}$ is a Gaussian:
 \begin{equation}
     W_{\ket{\beta}}(\alpha) = \frac{2}{\pi}e^{-2|\beta - \alpha|^2}.
     \label{eq:coherent-wigner}
 \end{equation}
In fact, coherent states are themselves part of a larger family known as Gaussian states. 
By virtue of Hudson's theorem, a pure state is `Gaussian' (has a Gaussian Wigner function) if and only if its Wigner function is non-negative \cite{GaussianQI-Weedbrook}.
As a result, Gaussian states can be classically simulated efficiently \cite{positive_wigner_sim-Eisert, Veitch_2013}. 
Whilst we will not go into details here (the interested reader can see App.~\ref{sect:sign-problem} and provided references), it suffices to mention that the positivity of the Wigner function allows one to treat it as a bona fide probability distribution, for which classical random sampling algorithms can be employed to output measurement statistics efficiently, polynomially scaling in the number of modes (provided the measurements are also Gaussian -- see below).

Gaussian operations (including measurements) are defined as those which map Gaussian states to Gaussian states, which can be implemented efficiently, in time $O(poly(m))$ for $m$ modes \cite{brask2022gaussian}. This is achieved by representing the state by its Wigner function, which for pure Gaussian states, is a Gaussian function, thus depending only on $O(m^2)$ numbers, to specify the first two moments of the distribution (means and variances of the quadratures). Gaussian operations include displacement operators, beamsplitters, phase shifts, as well as measurements of the position or momentum operators (homodyne)\footnote{Whilst the position and momentum eigenstates are not genuine quantum states, arbitrarily accurate Gaussian approximations can be found.}, or measuring in the coherent state basis (heterodyne).

\begin{figure}
    \centering
    \includegraphics[width=0.6\columnwidth]{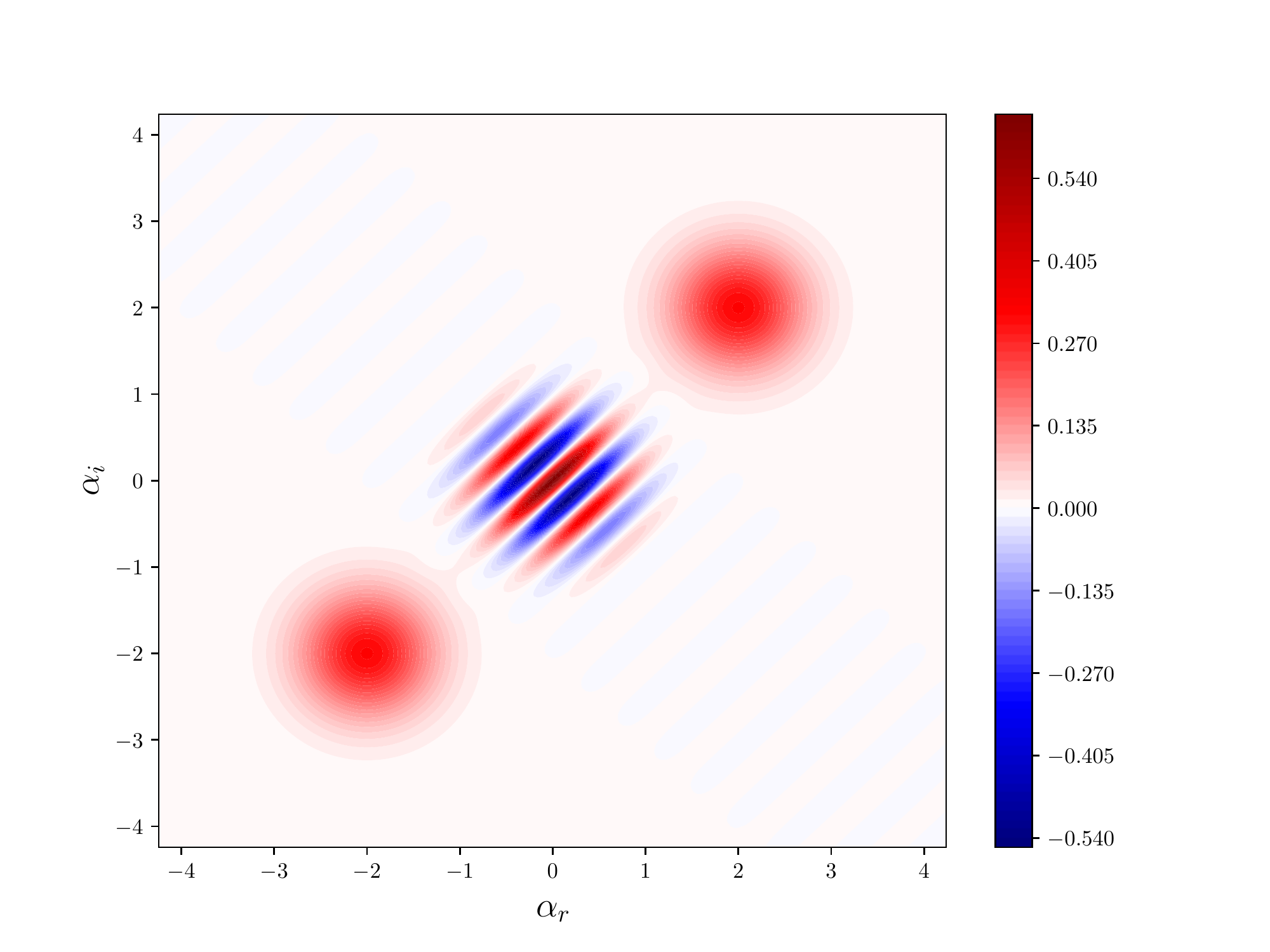}
    \caption{\textbf{Wigner function for even cat state.} The Wigner function, evaluated at $\alpha = \alpha_r + i\alpha_i$ for a normalized even cat state $\ket{2+2j}+\ket{-2-2j}$.}
    \label{fig:wf-cat}
\end{figure}

We can see an example where QPD positivity does not hold. The simplest case, perhaps, is a cat state, which is a normalized state of the form $\ket{\psi_\pm} = c(|\alpha\rangle \pm |-\alpha\rangle)$, where the $+$ ($-$) sign is often called an even (odd) cat state, due to the fact in the Fock basis, it only contains even (odd) photon numbers. Due to the non-orthogonality, the normalization condition $\braket{\psi_\pm}{\psi_\pm}=1$ is non-trivial, from computing overlaps via Eq.~\eqref{eq:coherent-overlap}: $2|c|^2 = 1/(1 \pm e^{-2|\alpha|^2})$. In App.~\ref{sect:derviation-off-diagonal-wigner} we show that the Wigner function for an off-diagonal term $\ketbra{\alpha}{\beta}$ is
\begin{equation}
    W_{\ketbra{\alpha}{\beta}}(\kappa) = e^{i \phi_{\alpha, \beta}(\kappa)}W_{\ket{\frac{1}{2}(\alpha+\beta)}}(\kappa).
    \label{eq:wigner-off-diagonal}
\end{equation}
That is, it is the Wigner function for the coherent state $\ket{\frac{1}{2}(\alpha+\beta)}$ (of the form Eq.~\eqref{eq:coherent-wigner}), multiplied by a $\kappa$ dependent phase factor. For a cat state $\ket{\psi_\pm}$, taking the normalization $c=|c|$, the Wigner function is therefore
\begin{equation}
    W_{\pm}(\kappa) = |c|^2(W_{\ket{\alpha}}(\kappa) +  W_{\ket{-\alpha}}(\kappa) \pm 2W_{\ket{0}}(\kappa)\cos [\phi_{\alpha, -\alpha}(\kappa)]),
    \label{eq:cat-wigner}
\end{equation}
where $\phi_{\alpha, -\alpha}(\kappa) = 4(\kappa_r \alpha_i - \kappa_i \alpha_r)$, with $\kappa_r = \mathrm{Re}(\kappa)$ etc.
Since the first two terms in Eq.~\eqref{eq:cat-wigner} are non-negative, all negativity is related to the phase $\phi$. In the limit of large $|\alpha|$, the Wigner function is effectively the sum of three independent functions; Gaussians at $\pm \alpha$, and a kind of modulated Gaussian at the origin. This is shown in Fig.~\ref{fig:wf-cat}, where the central oscillations (from the cosine term) are clearly observed. By Eq.~\eqref{eq:wigner-off-diagonal}, this negativity is due entirely to the off-diagonal `coherences' in the density matrix, $\ketbra{\alpha}{\beta}$.
Ref.~\cite{bourassa_fast_2021} provides a more general result of this discussion, allowing one to compute the Wigner function for `off-diagonal' Gaussian states.

We also briefly mention for completeness, that the theory of efficient simulation can be generalized beyond the Wigner function, for the Wigner function is just one QPD in a continuum \cite{QPD-sim_Caves}.
In all of these cases, the efficient QPD simulation is only possible when the relevant QPDs (states, operators, measurements) represent a genuine probability distribution, otherwise a sign problem of sorts is present, rendering the efficient classical simulation by these means in general infeasible. 
Indeed, QPD negativity is often considered a quantum resource, and as discussed in more detail in App.~\ref{sect:sign-problem}, the classical simulation complexity can be shown to scale exponentially in the number $n$ of non-Gaussian initial states, $\mathcal{N}^{2n}$, where $\mathcal{N}=\int d\alpha |W(\alpha)|\ge 1$ measures the amount of negativity ($\mathcal{N}=1$ iff the Wigner function is non-negative).

\subsubsection{Linear combination of Gaussians \label{sect:linear-gaussians}}

Refs.~\cite{bourassa_fast_2021, yao_recursive_2022} generalize the discussion above to allow simulations where the state is represented as a linear combination of Gaussian Wigner functions, $W_\rho(\alpha) = \sum_{g} c_g W_g(\alpha)$. In particular, it is demonstrated in Ref.~\cite{bourassa_fast_2021} how to represent many states of interest in this framework, including GKP and Fock states, the latter of which can describe $\ket{n}$ with $n+1$ Gaussian Wigner functions (to arbitrary accuracy). Update rules for Gaussian operations are given in \cite{yao_recursive_2022}, and non-Gaussian measurements can be performed by representing the measurement outcome as a linear combination of Gaussians, meaning that (Gaussian) Boson sampling can be simulated with time complexity scaling as ($O(2^n)$) $O(4^n)$, for $n$ photons in the output.

\subsection{Stellar rank \label{sect:stellar-sim}}
States in the Hilbert space can be constructed using the stellar representation \cite{bargmann, segal1963mathematical}. In particular, every single mode state $\ket{\psi} = \sum_{n=0}^\infty \psi_n \ket{n}$ has an associated stellar function 
\begin{equation}
    F_\psi(\alpha) = e^{\frac{1}{2}|\alpha|^2}\braket{\bar{\alpha}}{\psi} = \sum_{n=0}^\infty \psi_n \frac{\alpha^n}{\sqrt{n!}},
    \label{eq:stellar-func}
\end{equation}
which has the property $F_\psi(\hat{a}^\dag)\ket{0}=\ket{\psi}$.
The stellar rank $r(\psi)$ of a state is defined as the number of zeroes (counted with multiplicity) of $F_\psi$ (the stellar function is non-zero analytic on the complex plane), which is related to zeroes of the Husimi $Q$-function, $Q_\psi(\alpha) = e^{-|\alpha|^2} |F_\psi(\bar{\alpha})|^2/\pi$. For coherent states (and Gaussian states in general) the stellar rank is 0, whereas a state of at most $n$ photons, has stellar rank $n$.
A remarkable result from Ref.~\cite{stellar-rep-non-gaussian} is that any single mode state with finite stellar rank $r$, where the roots of the stellar function are $\{\bar{\beta}_i\}_{i=1}^r$, can be written as
\begin{equation}
    \ket{\psi} = \frac{1}{\sqrt{\mathcal{N}}} \left[\prod_{i=1}^r \hat{D}(\beta_i) \hat{a}^\dag \hat{D}^\dag(\beta_i) \right]\ket{G_\psi},
    \label{eq:psi-stellar-rep}
\end{equation}
where $\mathcal{N}$ is for normalization, and $\ket{G_\psi}=\hat{S}(\zeta_\psi)\ket{\alpha_\psi}$ a Gaussian state.

An additional and equally remarkable result is that the classical simulation complexity is intimately related to the stellar rank \cite{stellar-sim}.
% here I stress 'fock state' (as opposed to a general state) since otherwise the 's' parameter may be non-trivial, which is cumbersome to explain here.
In particular, a Fock state with stellar rank $r$ can be simulated under Gaussian operations/measurements with complexity $O(r^3 2^r + poly(m))$, for $m$ modes. Note for an initial state of $n$ photons (e.g. $\ket{1}^{\otimes n}$), the stellar rank is $r=n$, as the stellar rank is additive under tensor products, for pure states.

If one additionally wants to perform a measurement in a non-Gaussian basis (such as the Fock basis), this can be done at an additional cost related to the stellar rank of the output measurement, $r = r_\psi + \sum_{k=1}^m r_k$, where $r_k$ is the stellar rank for a particular output result in the $k$'th mode, and $r_\psi$ the stellar rank of the initial state \cite{ulysse-prl-boson-sampling}. For an $n$ photon simulation (such as Boson sampling), $r = 2n$, as the input and output each have rank $n$.

\subsection{Direct Fock state simulation of linear optics \label{sect:fock-sim}}
To simulate systems of strictly finite photon number there exists another, more `direct' approach, by simulating in the Fock basis.
Simulating Fock states is notoriously tricky, in fact this forms the basis for the classical hardness of Boson sampling \cite{aaronson-boson-sampling}. 
The general set up of interest in the present work is that of linear optics, where $n$ Fock states are injected over $m$ spatial modes (which are physically e.g., fibre optics or wave guides). That is, the initial state is of the form $\ket{n_1, n_2, \dots, n_n}\ket{0}^{\otimes (m-n)}$. The typical use case is for single photon states, i.e. $n_i=1$.
Linear optical (LO) operations -- beamsplitters and phase shifts -- are then applied between and on these spatial modes, after which Fock basis measurements are performed.
The best classical algorithms known for computing single amplitudes \cite{aaronson-boson-sampling, aaronson2013bosonsampling} and outputting measurement results \cite{clifford-clifford-2017} run in time (space) $O(n 2^n)$ ($O(m)$) for $n$ single photon inputs, based on calculations of the permanent (or slightly better in some cases \cite{clifford-clifford-2020}).
Full Fock state simulation of LO is however much more expensive in general, due to the size of the Hilbert space; for $n$ photons distributed amongst $m$ modes, the dimension is 
\begin{equation}
 d_{n,m} = {n+m-1 \choose m-1} = {n+m-1 \choose n},
 \label{eq:fock-dim}
\end{equation}
which can be significantly larger than $2^n$.

A general LO transformation is generated by a `bilinear' Hamiltonian, of the form
\begin{equation}
    \hat{H} = \sum_{i,j=1}^m h_{i,j} \hat{a}_i^\dag \hat{a}_j = \hat{H}^\dag,
    \label{eq:LO_ham}
\end{equation}
where the resulting unitary, $\hat{U}=e^{i\theta \hat{H}}$, has the property that it preserves the photon number, and therefore maps creation operators to a linear combination of creation operators \cite{kok_linear_2007}:
\begin{equation}
    \hat{a}_i^\dag \rightarrow \hat{U} \hat{a}_i^\dag \hat{U}^\dag = \sum_{j=1}^m u_{j,i} \hat{a}_j^\dag.
    \label{eq:operator-evolve}
\end{equation}
Here $u$ is an $m\times m$ unitary matrix, often called the `transfer matrix', describing the Heisenberg evolution of the operator\footnote{As such, the group of linear optical unitaries (also known as `passive transformations') is isomorphic to the unitary group of dimension $m$, $U(m)$.}.
That is, this tells us how a single photon propagates through a LO circuit. It is worth mentioning it is efficient, $O((N+m) m)$, to construct the transfer matrix $u$ of an entire LO circuit composed $N$ single or two-mode components (e.g., phase shifts, beamsplitters). 

Since such transformations preserve photon number, the evolution of a state can be described by evolving its creation operators:
\begin{equation}
    \hat{U}\ket{n_1, \dots, n_m} =\frac{1}{\sqrt{\mathcal{N}}} \hat{U} \left(\prod_{i=1}^n \hat{a}_{m_i}^\dag \right) \hat{U}^\dag \hat{U}\ket{0, \dots, 0} = \frac{1}{\sqrt{\mathcal{N}}}  \prod_{i=1}^n \left( \hat{U}\hat{a}_{m_i}^\dag \hat{U}^\dag\right) \ket{0, \dots, 0}.
\end{equation}
Here in the first step, we write each $\ket{n_i}=\frac{1}{\sqrt{n_i!}} (\hat{a}^\dag)^{n_i}\ket{0}$, and we inserted the identity operator $\hat{U}^\dag \hat{U}$ ($\mathcal{N}$ represents the normalization of the $n=\sum_i n_i$ photon state, and $\hat{a}_{m_i}^\dag$ creates a photon in the $m_i$'th mode). Then using the form of $\hat{H}$, we note $\hat{U}\ket{0, \dots, 0} = \ket{0, \dots,0}$, and we again insert the identity between each creation operator in the product, to arrive at the final form.
In particular then, the output can be computed by repeatedly applying Eq.~\eqref{eq:operator-evolve} on the operator representation:
\begin{equation}
    \prod_{i=1}^n \hat{a}^\dag_{m_i} \rightarrow \prod_{i=1}^n \left(\sum_{j=1}^m u_{j,m_i}\hat{a}_j^\dag \right).
    \label{eq:fock-evolution}
\end{equation}
By multiplying together all of these size $m$ polynomials (computing all Fock basis amplitudes), in the worst case, results in $d_{n,m}$ complex amplitudes to store in memory.
Since at step $i$ in the above multiplication, the current dimension is at most $d_{i,m}$, and this `state' is then multiplied by $m$ terms, the worst case time complexity for computing the output is $m\sum_{i=1}^{n-1} d_{i,m} = O(n d_{n,m})$ (the worst case memory requirements are $O(d_{n,m}$)). Sampling from the distribution therefore has a time complexity also scaling as $O(n d_{n,m})$.

Of course, in systems with symmetry, not every mode will be populated, and the actual cost can be significantly less (the simulation cost is linear in the number of Fock states with non-zero amplitude).
One can review Ref.~\cite{heurtel_strong_2023} for a more detailed expos\'e on the current state of the art for Fock simulation.

In the next section, we will outline a simulation approach using coherent states, where the simulation cost can be much better than the full Fock dimension, even if all Fock states are populated. In particular, whilst the simulation cost, in the worse case, is exponential in the number of photons, it is only quadratic in the number of modes.

\section{Simulation by coherent state rank \label{sect:sim}}
Here we outline a general strategy for simulating infinite dimensional systems using a finite rank decomposition over coherent states, i.e., states of the form $\sum_{i=1}^k c_i\ket{\alpha_i^{(1)}, \dots, \alpha_i^{(m)}}$ for $m$ modes, where each $\ket{\alpha_i^{(j)}}$ is a coherent state, and $k$ is the `coherent rank'.
In general we will allow for such a decomposition to be approximate, in the following sense:

\begin{definition}
The `approximate coherent rank' of a state $\ket{\psi}$ is the smallest integer $k$ such that for any $\epsilon>0$, there exists a coherent rank $k$ state $\ket{\tilde{\psi}}$ where $|\braket{\psi}{\tilde{\psi}}|^2 > 1 - \epsilon$.
\label{def:approx-coherent-rank}
\end{definition}

With such a representation, the memory requirements are simply from storing $k(m+1)$ complex numbers, i.e. each of the $c_i$, and the $\alpha_i^{(j)}$.
Overall we follow a similar approach as in finite rank stabilizer simulations, see e.g. \cite{Bravyi2019simulationofquantum}. In particular, we first show how to decompose Fock states into an approximate finite rank coherent state representation. We then discuss operations that are `free' within this approach, i.e., those that do not increase the rank, and can be applied in time $O(k)$. Following this we consider `resourceful' operations (that increase the rank), and then various methods to perform Fock basis measurements.

We will see in many situations this will outperform the direct Fock evolution (Eq.~\eqref{eq:fock-evolution}), potentially exponentially so, as the time complexity for computing amplitudes scales similarly as methods relying upon the permanent.

\subsection{Fock basis decomposition}
To simulate a Fock state using coherent states, we must first outline a strategy for converting a (possibly multi-mode) Fock state to a coherent state superposition. This relies upon the following:
\begin{theorem}
A single mode state in the Fock basis containing at most $N$ photons, $\ket{\psi}=\sum_{n=0}^Na_n \ket{n}$, has approximate coherent rank at most $N+1$. One explicit construction for this is
\begin{equation}
\boxed{
\begin{split}
        & \ket{\tilde{\psi}}= \frac{1}{\sqrt{\mathcal{N}}} \sum_{k=0}^N c_k \ket{\epsilon e^{2\pi i k / (N+1)}} \\
        &  c_k = \frac{e^{\epsilon^2/2}}{N+1}\sum_{n=0}^N \sqrt{n!} \frac{a_n}{\epsilon^n}e^{-2\pi i n k/(N+1)}
\end{split}
}
\label{eq:fock-fourier}
\end{equation}
where the normalization factor $\mathcal{N} = 1 + O(\epsilon^{2(N+1)}/(N+1)!)$  determines the fidelity $|\langle \psi|\tilde{\psi}\rangle|^2 = 1/\mathcal{N}$, and $\epsilon\in \mathbb{R}$ is a free parameter.
\label{th:general-fock}
\end{theorem}

\textit{Proof.} We show this by construction. Consider the target state $\ket{\psi}=\sum_{n=0}^N a_n\ket{n}$, where $\sum_n |a_n|^2=1$ (the protocol can also be applied to unnormalized `states' --- one can always first normalize it and later multiply be the inverse -- but for ease of exposition, we assume it is normalized here). Using Eq.~\eqref{eq:coherent-state}, we wish to satisfy the relations
\begin{equation*}
    a_n = \sum_{k=0}^N c_k e^{-\frac{1}{2}|\alpha_k|^2} \frac{\alpha_k^n}{\sqrt{n!}}
\end{equation*}
for $n=0, \dots, N$, where the approximate state is $\ket{\tilde{\psi}}=\sum_{k=0}^N c_k \ket{\alpha_k}$. We can achieve this by Fourier analysis. We simply posit $\alpha_k = \epsilon e^{2\pi i k/(N+1)}$, where for now $\epsilon \in \mathbb{R}$ is a free parameter (in practice $\epsilon$ could be complex, though it would not meaningfully change anything). With this, the equations become
\begin{equation*}
    \sum_{k=0}^N c_k e^{2\pi i n k/(N+1)} = e^{\frac{1}{2}\epsilon^2}\frac{a_n \sqrt{n!}}{\epsilon^n}.
\end{equation*}
Notice this can be solved by setting
\begin{equation}
    c_k = \frac{e^{\epsilon^2/2}}{N+1}\sum_{n=0}^N \sqrt{n!} \frac{a_n}{\epsilon^n}e^{-2\pi i n k/(N+1)},
    \label{eq:fourier_coeffs}
\end{equation}
using the properties of the roots of unity:
\begin{equation}
    \sum_{k=0}^{n-1} e^{2\pi i m k / n} = \left\{\begin{array}{cc}
        n & m/n \in \mathbb{N} \\
         0  &  \mathrm{otherwise}.
    \end{array} \right.
    \label{eq:unity-sim}
\end{equation}

With this, Eq.~\eqref{eq:fourier_coeffs} will exactly reproduce the amplitudes on the Fock states $\ket{0},\dots, \ket{N}$, however, in order for this to be of use, we must bound the error (e.g. the infidelity of the target state with respect to the constructed state). One can compute the amplitude on the next few Fock states in the expansion of $\ket{\tilde{\psi}}$:
\begin{equation}
    \langle N+m|\tilde{\psi}\rangle = e^{-\epsilon^2/2}\sum_{k=0}^N c_k \frac{\epsilon^{N+m}}{\sqrt{(N+m)!}} e^{2\pi ik (N+m)/(N+1)} = \epsilon^{N+1}a_{m-1}\sqrt{\frac{(m-1)!}{{(N+m)!}}},
\end{equation}
where here we restrict, for simplicity, $m=1, \dots, N+1$ (for $m>N+1$ a similar result holds, but we'd need to work modulo $N+1$).
Here the last step simply inserts the sum over $n$ from Eq.~\eqref{eq:fourier_coeffs}, and using Eq.~\eqref{eq:unity-sim} it picks out $n=m-1$ in the sum.
To compute the infidelity with respect to the target state, we must first normalize $\ket{\tilde{\psi}}$, which we do by noting the norm squared is
\begin{equation}
    \langle\tilde{\psi}|\tilde{\psi}\rangle = 1 + O\left(\frac{\epsilon^{2(N+1)}}{{(N+1)!}}\right),
\end{equation}
by the above result (the `1' comes simply from the fact the target state is normalized, and the approximation perfectly overlaps on those amplitudes, by construction).
By appropriately normalizing $\ket{\tilde{\psi}}$, we see, for small enough $\epsilon$:
\begin{equation}
    1 - |\langle \psi|\tilde{\psi}\rangle|^2 = O\left(\frac{\epsilon^{2(N+1)}}{{(N+1)!}}\right).
\end{equation}
This completes the proof, as $\epsilon$ is a free parameter, that can be used to tune the error arbitrarily small. \hfill \qedsymbol

This leads us to the following corollary:
\begin{corollary}
The state 
\begin{equation}
    \ket{\tilde{N}} = \frac{1}{\sqrt{\mathcal{N}}}\frac{\sqrt{N!}}{N+1}\frac{e^{\epsilon^2/2}}{\epsilon^N}\sum_{k=0}^N e^{-2\pi i k N/(N+1)}\ket{\epsilon e^{2\pi i k / (N+1)}}
    \label{eq:fock_approx}
\end{equation}
approximates the Fock state $\ket{N}$ with fidelity 
\begin{equation*}
|\langle N|\tilde{N}\rangle|^2 = \frac{1}{\mathcal{N}} = 1 - O\left(\frac{N!}{(2N+1)!}\epsilon^{2(N+1)}\right),
\end{equation*}
where, for normalization, $\mathcal{N} = N! \sum_{k=0}^\infty \frac{\epsilon^{2k(N+1)}}{(k(N+1)+N)!}$.
\label{cor:fock}
\end{corollary}
\textit{Proof.} This follows immediately from direct application of Eq.~\eqref{eq:fock-fourier}. \hfill \qedsymbol

We can observe that for a single photon state, $N=1$, this is an odd (normalized) cat state of the form $\ket{\epsilon} - \ket{-\epsilon}$, where one can see by inspection that this indeed tends to $\ket{1}$ as $\epsilon \rightarrow 0$ (all even Fock terms are 0 be construction, and all weight will concentrate on the $\ket{1}$ state for small enough $\epsilon$). For higher photon number Fock states this idea generalizes; we see that $\ket{\tilde{N}}$ only has non-zero amplitude on $\ket{N}, \ket{2N+1}, \ket{3N+2},\dots$.

After submitting the first version of this manuscript, we became aware that similar results as Cor.~\ref{cor:fock} have previously been observed. For example, in Ref.~\cite{grimsmo_quantum_2020} a projector onto select Fock states is constructed as a sum of $N+1$ phase shifts, $\sum_{k=0}^N  e^{(\hat{n}-N) 2 \pi i k /(N+1)}$, which when acting on a coherent state $\ket{\epsilon}$ gives Eq.~\eqref{eq:fock_approx}, upon normalization. Additionally,  decompositions of this type were also used in Refs.~\cite{vogel_unified-quantification, chabaud_quantum-inspired_2022}.

This shows, to approximate a Fock state of the form $\ket{n_1, n_1, \dots, n_m}$ results in a superposition of $\prod_{i=1}^m (n_i+1)$ coherent states. For example, the initial state for Boson sampling, $\ket{1}^{\otimes n}\ket{0}^{\otimes (m-n)}$ results in a superposition of $2^n$ coherent states. We will see, that under certain `free operations' that do not increase the rank (Sect.~\ref{sect:free-ops}), this can be much more efficient than storing each individual Fock amplitude, as was discussed in Sect.~\ref{sect:fock-sim}. Moreover, a simulation of a system with initial state $\ket{n}\ket{0}^{\otimes (m-1)}$ is only a superposition of $n+1$ coherent states. Thus, under free operations, such a system is efficient to simulate, whereas in the Fock basis this could result in $d_{n,m}$ (Eq.~\eqref{eq:fock-dim}) amplitudes to store.

We also mention here that Th.~\ref{th:general-fock} is technically only an upper bound on the approximate coherent rank, since we do not rule out more efficient representations (i.e., the optimality of Th.~\ref{th:general-fock} is an open question). It is however easy to verify for the Fock state $\ket{1}$, the approximate coherent rank is \textit{exactly} 2.

\begin{figure}
    \centering
    \includegraphics[width=0.7\columnwidth]{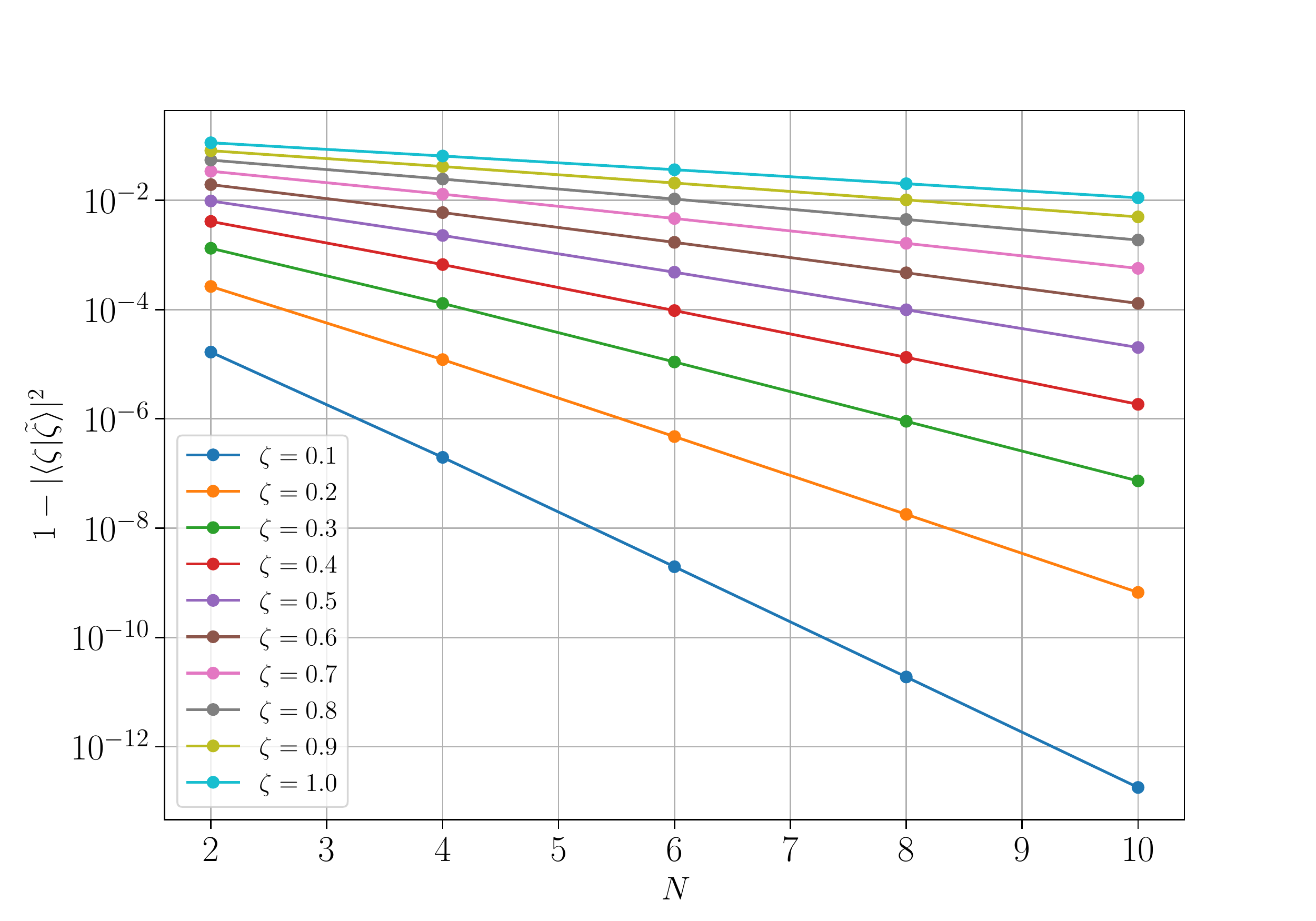}
    \caption{\textbf{Squeezed vacuum state approximation.} Infidelity of approximate squeezed vacuum state $\ket{\tilde{\zeta}}$ with respect to reference target $\ket{\zeta}$ of Eq.~\eqref{eq:squeezed_state} (note, here we set $\phi=0$ in $\zeta=re^{i\phi}$). The approximation $\ket{\tilde{\zeta}}$ is a superposition of $N$ coherent states as described in App.~\ref{sect:squeeze-approx}.}
    \label{fig:squeeze_approx}
\end{figure}

Whilst the above construction was presented for finite Fock state representations (i.e. up to $N$ photons), it can, in certain cases, be used to approximate states with support over the entire Hilbert space. As a pertinent example, we apply this to squeezed vacuum states $\ket{\zeta}$ (as used in Gaussian Boson sampling \cite{GBS-Hamilton}), which by Eq.~\eqref{eq:squeeze-operator} is:
\begin{equation}
    \ket{\zeta} = \hat{S}(\zeta)\ket{0} = \frac{1}{\sqrt{\cosh r}}e^{-\frac{1}{2}e^{i\phi} \tanh (r) \hat{a}^{\dag 2}} \ket{0} = \frac{1}{\sqrt{\cosh r}} \sum_{n=0}^\infty (-e^{i\phi}\tanh r)^n \frac{\sqrt{(2n)!}}{2^n n!}\ket{2n},
    \label{eq:squeezed_state}
\end{equation}
where $\zeta = re^{i\phi}$.
Since the amplitudes tend to zero as $n$ grows (albeit slowly for large $r$)\footnote{The ratio of amplitudes $|\braket{2n+2}{\zeta}|/|\braket{2n}{\zeta}| = \tanh r \sqrt{(2n+1)/(2n+2)}  < 1, \forall r$.}, it is possible to approximate such a state to arbitrary accuracy, by only taking the first $N$ terms, for some $N=N(r)$. In Fig.~\ref{fig:squeeze_approx} we show such an approximation.
Indeed, to achieve a particular target fidelity depends strongly on the squeezing parameter $r$. Nevertheless, this demonstrates such a procedure can be used in principle.
 Note, the approximation we use here does not follow Eq.~\eqref{eq:fock-fourier} precisely (though the general idea is the same), but a related form to take advantage of the structure of Eq.~\eqref{eq:squeezed_state}, as shown in App.~\ref{sect:squeeze-approx}.

Lastly, we show we can rigorously upper bound the coherent state rank, in particular:
\begin{proposition}
 A single mode state $\ket{\psi}$ with finite stellar rank $r$,
\begin{equation}
\ket{\psi} = \frac{1}{\sqrt{\mathcal{N}}} \left[\prod_{i=1}^r \hat{D}(\beta_i) \hat{a}^\dag \hat{D}^\dag(\beta_i) \right]\ket{G_\psi},
\label{eq:psi-stellar}
\end{equation}
has approximate coherent rank at most $k (r+1)$, where $k$ is the coherent rank of $\ket{G_\psi}$.
\label{prop:stellar-coherent-ranks}
\end{proposition}
\textit{Proof.} Using Eq.~\eqref{eq:displacement-creation-annihilation} we have,
\begin{align}
\prod\limits_{j=1}^{r} \hat{D}(\beta_{j}) \hat{a}^{\dagger} \hat{D}^{\dagger}(\beta_{j}) = \prod\limits_{j=1}^{r} \left( \hat{a}^{\dagger} - \overline{\beta}_{j} \right).
\end{align}
Therefore, the operator \(\prod\limits_{j=1}^{r} \hat{D}(\beta_{j}) \hat{a}^{\dagger} \hat{D}^{\dagger}(\beta_{j}) \) contains \textit{at most} \(\left( a^{\dagger} \right)^{r}\) in the expansion, and can be written in the general form \( \sum_{l=0}^r b_l \left( \hat{a}^{\dagger} \right)^{l}\). Let \(| G_{\psi} \rangle\) be a pure state with coherent rank \(k\), that is, \(| G_{\psi} \rangle = \sum_{j=1}^{k} c_{j} | \alpha_{j} \rangle\). Then, Cor.~\ref{cor:creation_cost} below shows that applying a resourceful operation of the form \( \sum_{l=0}^r b_l \left( \hat{a}^{\dagger} \right)^{l}\) on a coherent rank-\(k\) state increases its approximate coherent rank to \(k(r+1)\). \hfill \qedsymbol

It is important to note that, in the stellar rank formalism, the state \(| G_{\psi} \rangle\) is a general Gaussian state and therefore, it may have nontrivial squeezing \(\zeta \neq 0\), which can result in an arbitrarily large coherent rank\footnote{For large $|\zeta|$, the amplitude $|\braket{2n}{\zeta}| \approx \sqrt{\frac{2}{e^{|\zeta|}}} \frac{\sqrt{(2n)!}}{2^{n} n!}$ becomes arbitrarily small.}. 
However, for $\zeta=O(1)$, by keeping terms up to \(n\)'th order in the Taylor expansion of \(\hat{S}(\zeta)\), we can approximate \(| \zeta \rangle\) with approximate coherent rank \(2n+1\).

\subsection{Free operations \label{sect:free-ops}}
We consider the class of operations that do not increase the rank, $k$, for states of the form $\sum_{i=1}^k c_i\ket{\alpha_i^{(1)}, \dots, \alpha_i^{(m)}}$. Generally speaking, such operations are those that map  a tensor product of $m$ coherent states into another tensor product of $m$ coherent states.
In particular such operations generate no mode entanglement when acting on coherent states, which contains the set of linear optical operations.
In line with the resource theory literature, we call such operations `free operations'.

The first obvious example is a single mode displacement operator, which to update the state, takes time $O(k)$, simply by updating the relevant single-mode coherent state for each term in the sum, as well as the amplitudes: $\hat{D}(\beta)\ket{\alpha} = e^{i \mathrm{Im}( \bar{\alpha}\beta)}\ket{\alpha + \beta}$. 

Another obvious and relevant example is a single-mode phase shift, as used in linear optics, $\hat{P}(\phi) = e^{i\phi \hat{n}}$. It is easy to see such an operation acts as $\hat{P}(\phi)\ket{\alpha} = \ket{e^{i\phi}\alpha}$.

In fact, the result can be generalized to that of \textit{all} LO operations:

\begin{theorem}
    All linear optical unitary transformations $\hat{U}$, with respective transfer matrix $u$ (Eq.~\eqref{eq:operator-evolve}), are free. The time complexity for updating a rank $k$ state is $O(k\ell^2)$, where $\hat{U}$($u$) acts non-trivially on $\ell \le m$ modes.
\label{th:lo-free}
\end{theorem}
\textit{Proof.} Consider the action of such a unitary on $\ket{\alpha_1, \dots, \alpha_m}$:
\begin{equation*}
    \hat{U}\ket{\alpha_1, \dots, \alpha_m} = \hat{U} \prod_{i=1}^m \hat{D}_i(\alpha_i) \ket{0, \dots, 0} =  \prod_{i=1}^m \left( \hat{U} \hat{D}_i(\alpha_i) \hat{U}^\dag\right) \hat{U}\ket{0, \dots, 0} = \prod_{i=1}^m \left( \hat{U} \hat{D}_i(\alpha_i) \hat{U}^\dag\right) \ket{0, \dots, 0},
\end{equation*}
where we insert the identity $\hat{U}\hat{U}^\dag = \mathbb{I}$ between each displacement operator ($\hat{D}_i$ only acting non-trivially on mode $i$), and the last step uses that for LO unitaries, $\hat{U}\ket{0,\dots, 0}=\ket{0, \dots, 0}$ (conservation of particle number).

Using that $\hat{U}e^{\hat{X}}\hat{U}^\dag = e^{\hat{U}\hat{X}\hat{U}^\dag}$ by unitarity, we can write
\begin{equation*}
    \hat{U}\hat{D}_i(\alpha)\hat{U}^\dag = \hat{U}e^{\alpha \hat{a}_i^\dag - \bar{\alpha}\hat{a}_i} \hat{U}^\dag = e^{\alpha \hat{U}\hat{a}_i^\dag\hat{U}^\dag - \bar{\alpha}\hat{U}\hat{a}_i\hat{U}^\dag}.
\end{equation*}
Now by the linear property, Eq.~\eqref{eq:operator-evolve}, we have
\begin{equation*}
    \hat{U}\hat{a}_i^\dag\hat{U}^\dag = \sum_{j=1}^m u_{j,i} \hat{a}_j^\dag;\;\;\hat{U}\hat{a}_i\hat{U}^\dag = \sum_{j=1}^m \bar{u}_{j,i} \hat{a}_j,
\end{equation*}
and therefore
\begin{equation*}
    \hat{U}\hat{D}_i(\alpha)\hat{U}^\dag = \exp \left[\sum_{j=1}^m (\alpha u_{j,i} \hat{a}_j^\dag - \bar{\alpha}\bar{u}_{j,i}\hat{a}_j)\right] = \prod_{j=1}^m \hat{D}_j(\alpha u_{j,i}).
\end{equation*}
That is, a single mode displacement operator is mapped to a multi-mode displacement operator, and combining these results we see the unitary $\hat{U}$ maps a multi-mode coherent state, to a multi-mode coherent state:
\begin{equation}
    \hat{U}\ket{\alpha_1, \dots, \alpha_m} = \prod_{i,j=1}^m \hat{D}_j(\alpha_i u_{j,i})\ket{0, \dots, 0} = \bigotimes_{j=1}^m \left| \left. \sum_{i=1}^m \alpha_i u_{j,i} \right\rangle \right..
    \label{eq:general-lo-evolve}
\end{equation}
The last step uses that a product of displacement operators is also a displacement operator (Eq.~\eqref{eq:displacement-family}), where the argument is summed. In general this generates a phase factor, but by unitarity of $u$ it is easy to verify this is 0 here; the phase  is composed entirely of terms of the form $\alpha_i \bar{\alpha}_j\sum_{k}u_{k,i}\bar{u}_{k,j}=0$ (with $i\neq j$).

For LO transformations therefore, the update can be performed by simple matrix-vector multiplication,
\begin{equation}
    \ket{\alpha_1', \dots, \alpha_m'} = \hat{U}\ket{\alpha_1, \dots, \alpha_m};\;\;     \left(\begin{array}{cc}
        \alpha_1' \\
        \vdots \\
        \alpha_m'
    \end{array}\right) = u \left(\begin{array}{cc}
        \alpha_1 \\
        \vdots \\
        \alpha_m
    \end{array}\right),
    \label{eq:general-lo-update}
\end{equation}
which in the (worst) case where $u$ acts non-trivially over all modes, has computational cost $O(m^2)$. 

The above shows the coherent rank of a state is unchanged by any LO operation.
Further, if $\hat{U}$ only acts non-trivially on $\ell \le m$ modes, then the relevant submatrix of $u$ has size $\ell \times \ell$, resulting in computational cost to update a rank $k$ state, $O(k \ell^2)$.
\hfill \qedsymbol

We can use Eq.~\eqref{eq:general-lo-update} to show for a beamsplitter described by a unitary $\hat{B}$ with transfer matrix $u$,
\begin{equation}
\hat{B}(\theta, \phi) = e^{\frac{\theta}{2}(\hat{a}^\dag \hat{b} e^{i\phi} - \hat{a} \hat{b}^\dag e^{-i\phi})}, \;\; u = \left(\begin{array}{cc}
        t & re^{i\phi} \\
        -re^{-i\phi} & t
    \end{array}\right),
\label{eq:beamsplitter}
\end{equation}
where $t=\cos \frac{\theta}{2}, r=\sin \frac{\theta}{2}$, that 
\begin{equation}
    \hat{B}(\theta, \phi)\ket{\alpha, \beta} = \ket{t\alpha + re^{i\phi}\beta, t\beta - re^{-i\phi}\alpha}.
    \label{eq:beamsplitter_action}
\end{equation}

Th.~\ref{th:lo-free} shows that \textit{all} linear optical operations are free within the framework of coherent state rank simulation. In particular, the time complexity to compute the final state (before measurement) under a general Boson sampling type (random) circuit of $m$ modes and $n$ photons is $O(m^2 2^n)$, and space complexity $O(m 2^n)$. This generally will be much less costly than the full simulation in the Fock basis (Sect.~\ref{sect:fock-sim}).

\begin{figure}
    \centering
    \includegraphics[width=0.98\columnwidth]{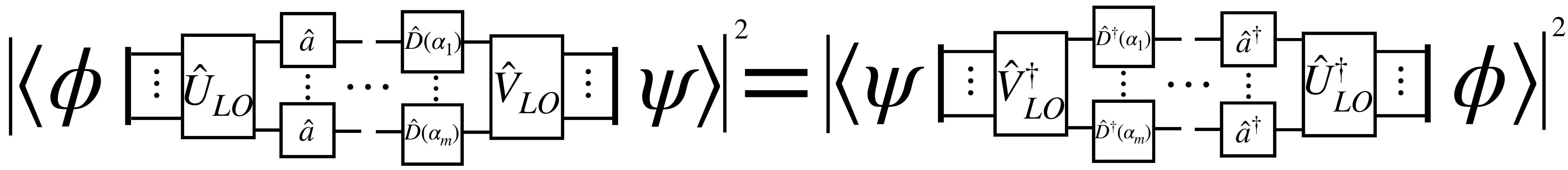}
    \caption{\textbf{Example of a free circuit, and its inverse}. The application of only linear optical (LO) unitaries, displacements, and annihilation operators are free, meaning the coherent rank of $\ket{\psi}$ is unchanged under such a circuit (left). Similarly if one has a circuit of only LO unitaries, displacements, and creation operators (right), the output probabilities can be computed with the same cost as the time-reversed (free) circuit. Circuits with both creation and annihilation operators however incur additional cost for implementing the non-free operations.}
    \label{fig:circuit-example}
\end{figure}

Lastly, whilst not a unitary operation, it is worth noting that, essentially by definition, the annihilation operator is also free\footnote{In fact, it is a free \textit{operation} as opposed to a free unitary, since it can consume resource, see App.~\ref{sec:free-unitaries}.}, as it just modifies each complex amplitude $c_i\rightarrow c_i\alpha_{i}^{(j)}$, when applying $\hat{a}_j$ (therefore taking time $O(k)$ to update the state). Additionally, any function $f(\hat{a})$ can be similarly implemented (assuming it can be computed classically in polynomial time). In general the creation operator is not free, as we will see below. However, noting that $ \langle \phi| \hat{a}|\psi\rangle  = \overline{\langle \psi |\hat{a}^\dag|\phi\rangle}$, the simulation cost for computing output amplitudes/probabilities in circuits composed of free unitaries and \textit{only} creation operators, is equivalent to the time-reversed circuit containing entirely free (unitary and annihilation) operators. See Fig.~\ref{fig:circuit-example} for an example.

\subsection{Resourceful operations}
Any operation that is not classed as free is `resourceful'. Such an operation, when acting on a multi-mode coherent state will generate a non-trivial superposition. One example is a single mode squeezing operator Eq.~\eqref{eq:squeeze-operator}, which when acting on the (coherent) vacuum generates a squeezed vacuum state, Eq.~\eqref{eq:squeezed_state}, and hence squeezing is not a free operation (see App.~\ref{sect:squeeze-approx} for one decomposition).
Another example (though not unitary) is the creation operator, $\hat{a}^\dag$, which is clearly not free. In fact, as we will see below in Cor.~\ref{cor:creation_cost}, $\hat{a}^\dag$ causes the rank of a state to double. We can first make a definition analogous to Def.~\ref{def:approx-coherent-rank} at the operator level:

\begin{definition}
    The (approximate) `operator coherent rank' for an $m$-mode operator $\hat{A}$ is $\ell$, if and only if $\hat{A}\ket{\vec{\alpha}}$ has (approximate) coherent state rank at most $\ell$, for all $\ket{\vec{\alpha}}$.
    \label{def:operator-rank}
\end{definition}

In other words, the action of an operator with coherent rank $\ell$ on a coherent state results in a state of coherent rank at most $\ell$ (and this is achieved for at least one coherent state).
One can readily observe that an operator with (approximate) operator coherent rank $\ell$ can be implemented on a state with (approximate) coherent rank $k$, resulting in a state of at most (approximate) coherent rank $k\ell$.

An operator composed as a sum of displacements, $\hat{U} = \sum_{i=1}^{{\ell}} a_i \hat{D}_i(\alpha_i)$, has operator coherent rank at most $\ell$, and similarly for an operator represented in the coherent state basis, $\hat{U}=\sum_{i,j=1}^{{\ell}} a_{i,j}\ketbra{\alpha_i}{\beta_i}$.

One potential construction for this is to use a similar Fourier decomposition as in Th.~\ref{th:general-fock}, but at the level of an operator. Consider an operator acting on at most $N$ photons: $\hat{A}=\sum_{n,m=0}^N A_{n,m} \ketbra{n}{m}$. This could result, for example, from the truncation of an operator to the maximum $N$ photon subspace. Then one can use the following Proposition:
\begin{proposition}
    The operator $\hat{A}=\sum_{n,m=0}^N A_{n,m} \ketbra{n}{m}$ in the Fock basis up to $N$ photons has approximate operator coherent rank at most $N+1$. One explicit representation is
    \begin{equation}
    \begin{split}
        & \tilde{\hat{A}} = \sum_{k,l=0}^N c_{k,l}\ketbra{\epsilon e^{2\pi i k / (N+1)}}{\epsilon e^{2\pi i l / (N+1)}}, \\
        & c_{k,l} = \frac{e^{|\epsilon|^2}}{(N+1)^2} \sum_{n,m=0}^N \sqrt{n!m!}\frac{A_{n,m}}{\epsilon^{n+m}} e^{-2\pi i (kn-lm)/(N+1)},
        \end{split}
    \end{equation}
    which is arbitrarily accurate for $\epsilon \rightarrow 0$.
\end{proposition}
\textit{Proof.} This follows almost exactly that of Th.~\ref{th:general-fock}. \hfill \qedsymbol

Such a decomposition can also be applied to multi-mode operators. One clear issue with such a method is that if $N$ is large, the state rank $k$ will increase very quickly $k\rightarrow k(N+1)$, meaning very few operators in this form can be applied. But nevertheless, this allows one, in principle, to achieve universality with displacements and SNAP operations $\exp(i\theta_n\ketbra{n}{n})=\ketbra{n}{n}(e^{i\theta_n}-1) + \mathbb{I}$ \cite{universal-snap}, which has approximate operator coherent rank at most $n+2$.

A second, perhaps more practical approach, is to recompute each coherent state in the superposition to a desired accuracy in the Fock basis, and then reexpress the state via Th.~\ref{th:general-fock}. For example, in the single mode case $\hat{U}\ket{\psi} = \sum_{i=1}^k c_i \hat{U}\ket{\alpha_i}$, and one can truncate each $\hat{U}\ket{\alpha_i} = \sum_{n\ge0} a_n \ket{n} \approx \sum_{n=0}^{N_i} a_n \ket{n}$ to the $N_i$ photon subspace (to desired accuracy), which can be (approximately) written as $N_i+1$ coherent states. In this example, the output state has coherent rank at most $k + \sum_{i=1}^k N_i$.

Lastly we show some results pertaining to the implementation of creation operators. We will see a creation operator can be implemented doubling the rank, or in general, an operator composed of $n$ such operators will increase the rank by a factor of $n+1$:

\begin{theorem}
    Any single-mode operator composed of only creation and annihilation operators, of the form $\hat{A}:=\sum_{i=1}^p b_i \prod_{j=1}^{l_i} (\hat{a}^{\dag})^{n_{i,j}} (\hat{a})^{m_{i,j}}$, where there are at most $n$ creation operators in total for each term in the sum (i.e., $n \ge \sum_{j=1}^{l_i} n_{i,j}, \forall i$), has approximate operator coherent rank at most $n+1$.
\label{th:creation_annihilation_cost}
\end{theorem}
\textit{Proof.} Such an operator acts on a coherent state as:
 \begin{equation*}
     \hat{A}\ket{\alpha} = \hat{A} \hat{D}(\alpha) \ket{0} = \hat{D}(\alpha)\hat{D}^\dag(\alpha)\hat{A} \hat{D}(\alpha)\ket{0} =: \hat{D}(\alpha)\hat{A}'(\alpha)\ket{0}.
 \end{equation*}
 Now, from Eqs.~\eqref{eq:displacement-family}, \eqref{eq:displacement-creation-annihilation}, by repeatedly inserting the identity operator $\hat{D}^\dag(\alpha)\hat{D}(\alpha)$, the operator $\hat{A}'$ can be written as 
 \begin{equation*}
     \hat{A}'(\alpha) = \sum_{i=1}^p b_i \prod_{j=1}^{l_i} (\hat{a}^{\dag} + \bar{\alpha})^{n_{i,j}}(\hat{a} + {\alpha})^{m_{i,j}}.
 \end{equation*}
  This shows that $\hat{A}'\ket{0}$ defines a (generally unnormalized) state in the Fock basis containing at most $n$ photons, i.e., it is of the form $\hat{A}'\ket{0} = \sum_{i=0}^n c_i \ket{i}$. 
Moreover, note that it is efficient to compute the amplitudes $c_i$ (with worst case time scaling as $O(pn^2)$).
  By Th.~\ref{th:general-fock}, such a state can be decomposed to have approximate coherent rank (at most) $n+1$.
  The result follows by noting the final application of $\hat{D}(\alpha)$ leaves the rank unchanged, as displacements are free. \hfill \qedsymbol

This leads immediately to the observations (we also give an alternative proof of this in App.~\ref{sect:alternate-proof-adag}, using post-selection):
\begin{corollary}
    The operator $\hat{a}^\dag$ has approximate operator coherent rank $2$, and an operator $\sum_{l=0}^n b_l (\hat{a}^{\dag})^l$ has approximate operator coherent rank at most $n+1$. The classical simulation of photon-added coherent states therefore has a cost linear in the number of photon additions.
\label{cor:creation_cost}
\end{corollary}

Th.~\ref{th:creation_annihilation_cost} allows one to implement operators generated by the creation/annihilation operators. For example, squeezing operations (Eq.~\eqref{eq:squeeze-operator}) can be implemented to arbitrary accuracy by expanding $\hat{S}(\zeta)$ to order $n$ (terms up to $(\bar{\zeta}\hat{a}^2 - \zeta \hat{a}^{\dag 2})^n$), which will increase the states coherent rank by a factor of at most $2n+1$.

In general, applying multiple resourceful operations in a circuit causes an exponential overhead for the classical simulation complexity. For example, $n$ independent applications of the creation operator (potentially on different modes) will generically result in an $O(2^n)$ increase in the coherent state rank, by Cor.~\ref{cor:creation_cost}. This is similar to the case of stabilizer/magic simulations, where only a logarithmic number of $T$ gates can be tolerated \cite{Bravyi2019simulationofquantum}.
Of course, in some situations this can be improved upon as demonstrated above, where (for example) applying $n$ creation operators sequentially on a single mode only increases the simulation cost linearly (instead of exponential) in $n$, meaning a polynomial (instead of logarithmic) number can be applied.

\subsection{Measurements \label{sect:measurements}}
The last ingredient in the coherent state rank simulation paradigm is that of measurements. Here we primarily consider Fock basis measurements, but will also mention properties of coherent state basis measurement. There are several ways in which to perform a measurement in this approach, each of which has different pros and cons.

For consistency, we define an $m$ mode state of rank $k$ as before: $\ket{\psi} = \sum_{i=1}^k c_i\ket{\alpha_i^{(1)}, \dots, \alpha_i^{(m)}}$. This could correspond to the state after evolving a Fock state through a linear optical network, for example. We will see that computing the probability for a particular Fock/coherent basis measurement can be achieved in a time linear in the rank $k$, but for actually outputting measurement results according to the desired distribution, there are additional complications.

\subsubsection{Computing individual amplitudes \label{sect:compute-amps}}
Before discussing probabilistic measurement (sampling), we first observe that the computation of individual amplitudes/probabilities can be performed in time $O(mk)$ by evaluating the overlaps
\begin{equation}
\begin{split}
    & \langle n_1, \dots, n_m|\psi\rangle = \sum_{i=1}^k c_i \prod_{j=1}^m e^{-\frac{1}{2}|\alpha_i^{(j)}|^2}\frac{(\alpha_i^{(j)})^{n_j}}{\sqrt{n_j!}} \\
    & \langle \beta_1, \dots, \beta_n|\psi\rangle = \sum_{i=1}^k c_i \prod_{j=1}^m e^{-\frac{1}{2}|\beta_j - \alpha_i^{(j)}|^2}e^{-i\mathrm{Im}(\beta_j \bar{\alpha}_i^{(i)})},
\end{split}
\label{eq:fock_coherent_amps}
\end{equation}
where $\ket{n_i}$ is a Fock state of $n_i$ photons, and $\ket{\beta_i}$ a coherent state. That is, one can compute the probability of a particular Fock/coherent basis measurement result in time $O(mk)$. The computation of the above overlaps is equivalent to applying free operations (annihilations or displacements respectively) on $\ket{\psi}$, followed by projecting to the vacuum state. 
 In fact, the above can be generalized to computing the overlaps of two arbitrary states generated by (say) LO unitaries $\hat{U}, \hat{V}$, e.g., $\langle \phi |\psi\rangle = \langle \vec{0}|\prod_{i}\hat{a}_{i} \hat{V}^\dag \hat{U} \prod_{j} \hat{a}_j^\dag|\vec{0}\rangle$. Noting annihilation operators are free, the computational cost to compute this overlap is $O(m^2k)$ (Th.~\ref{th:lo-free}), where $k$ is the coherent rank of of the initial state $\prod_j \hat{a}_j^\dag|\vec{0}\rangle$ (we omit potential normalization factors for simplicity).

We further note that certain output amplitudes can be computed much more efficiently in the setting of Boson sampling, by simulating the time-reversed circuit. For example, since \newline $\langle n, 0, \dots, 0| \hat{U}|1,\dots, 1\rangle =\overline{\langle 1, \dots, 1| \hat{U}^\dag |n, 0, \dots, 0\rangle }$, and the state $\ket{n, 0, \dots, 0}$ has rank $n+1$, the simulation time complexity to compute this output amplitude is $O(m^2 n)$, instead of $O(m^2 2^n)$ for the forward circuit. In general, the cost to compute some particular transition amplitude is determined by which of the input or output has the most efficient coherent rank representation: 
\begin{equation*}
 \left\langle \vec{0}\left| \left(\prod_{j \in \mathcal{J}_{out}} \hat{a}_{j} \right) \hat{U} \left(\prod_{i \in \mathcal{I}_{in}} \hat{a}^\dag_{i} \right) \right|\vec{0} \right\rangle 
= 
\overline{\left\langle \vec{0} \left| \left(\prod_{{i \in \mathcal{I}_{in}}} \hat{a}_{i} \right) \hat{U}^\dag \left( \prod_{j \in \mathcal{J}_{out}} \hat{a}^\dag_{j} \right) \right|\vec{0} \right \rangle } .    
\end{equation*}

Moreover, projecting a state onto a partial measurement result of $p<m$ modes (without normalization), e.g., $\langle n_1, \dots, n_p|\psi\rangle$ or $\langle \beta_1, \dots, \beta_p|\psi\rangle$, has time cost $O(pk)$, given a rank $k$ state (i.e. such an operation is `free'). 
Note however, computing the normalization of the Fock projected state is, in general, not linear in $k$ (on the other hand, we will see coherent state projection can be normalized with cost linear in $k$). 
First, we can notice that since coherent states are not orthogonal, the cost for computing the norm of an $m$-mode rank-$k$ `state', by evaluating all overlaps in $\| |\psi\rangle \|^2=\braket{\psi}{\psi}$, has computational cost $O(mk^2)$.
This can be improved upon, if the state is generated by a free unitary. For example, if $\ket{\psi} = \hat{U}\prod_{i} \hat{a}_i^\dag \ket{\vec{0}}$ is of rank $k$, where $\hat{U}$ is a LO unitary, then the norm squared of the projected state can be computed as (up to trivial normalization factors from the creation operators):
\begin{equation}
    \|\langle n_1, \dots, n_p|\psi \rangle \|^2 = \overbrace{\langle \vec{0}|\prod_i \hat{a}_i \hat{U}^\dag}^{\bra{\psi}} \prod_{j=1}^p \hat{a}_j^{\dag n_j} (\ketbra{0}{0})^{\otimes p} \prod_{\ell=1}^p \hat{a}_\ell^{n_\ell} \overbrace{\hat{U}\prod_i \hat{a}_i^\dag |\vec{0}\rangle}^{|\psi\rangle}. 
    \label{eq:partial-norm}
\end{equation}
Applying the operations from right to left will result in a state of coherent rank $k \prod_{j=1}^p (n_j+1)$, the overhead coming from the creation operators (Cor.~\ref{cor:creation_cost}). Therefore computing the above overlap takes time $O(m^2 k \prod_{j=1}^p (n_j+1))$. 
For small $p$ this is generally more efficient than the direct $k^2$-scaling calculation, e.g., if $p=1$ it's simply $O(n_1 m^2k)$.
In the worst case however, it is actually less efficient, for example if all $n_j=1$, then the cost is $O(m^2 k 2^{p}$). As such, we can generically upper bound the complexity of evaluating the norm of the $(m-p)$-mode partially Fock projected state as $O((m-p)k^2)$, by direct calculation of all overlaps.

To calculate the equivalent conditional coherent basis normalization of $\langle \beta_1, \dots, \beta_p|\psi\rangle$ via Eq.~\eqref{eq:partial-norm}, one replaces the $\hat{a}_j^{\dag n_j}, \hat{a}_\ell^{n_\ell}$ with displacement operators $\hat{D}(\beta_j), \hat{D}(-\beta_\ell)$, which as they are free, the cost of computing this norm is linear in the rank, $O(m^2 k)$.

\subsubsection{Metropolis sampling algorithm}
For sake of an example, consider the case of interest being Boson sampling-like simulations, with $n$ single photon inputs in $m$ modes. The coherent rank of such a state is $k=2^n$, and the Fock dimension is $d_{n,m}$ (Eq.~\eqref{eq:fock-dim}).
In the most `naive' approach to outputting measurement samples from the desired distribution, one can simply compute measurement probabilities $p_{\vec{n}}$ until some (random) threshold $r<1$ is reached, $\sum_{\vec{n}} p_{\vec{n}}>r$. Although computing each individual measurement probability is linear in $k$, $O(m k)$, this procedure may require the evaluation of $O(d_{n,m})$ probabilities, i.e., resulting in overall time complexity $O(m 2^n d_{n,m})$. This can therefore be more costly (in time) than doing the full Fock basis simulation, which scales as $O(n d_{n,m}$) for outputting measurement samples, as discussed in Sect.~\ref{sect:fock-sim} (the memory overhead will still in general be better in the coherent representation however, $O(m2^n)$ vs. $O(d_{n,m})$).

We can improve upon this `naive' sampling,  by a heuristic approach, using a Markov chain.
This could also be applied to the coherent state basis, though as it is a continuous distribution, we focus on Fock measurements here for simplicity.
Similar methods have been proposed and used for Boson sampling \cite{mcmc-bs-Liu_2020} and in stabilizer rank simulation \cite{Bravyi2019simulationofquantum}. We will outline a very high level version of this type of algorithm as an example for the reader, but mention that this can be implemented in many different ways, each likely with different benefits and drawbacks.

The main idea is to set up a Markov chain, where each state in the chain is a measurement outcome. One can propose new outcomes, and accept them with the appropriate probability. For example:
\begin{enumerate}
    \item Input: A rank-$k$ $m$-mode state $\ket{\psi}$
    \item Initialization: Pick a random Fock basis measurement outcome $\vec{n}=(n_1, \dots, n_m)$, and compute $P_{\vec{n}} = |\langle \vec{n}|\psi\rangle|^2$
    \item Propose a new measurement outcome $\vec{n'}$ by local moves from $\vec{n}$, and compute $P_{\vec{n'}}$
    \item Accept change with probability $\mathrm{min}\{1, P_{\vec{n'}}/P_{\vec{n}}\}$
    \item Repeat steps 3, 4 for $T$ steps, where $T$ is a user determined `thermalization' time, and output the current state
\end{enumerate}

Some comments are in order:
i) In step 2, the $n_i$ should be chosen to take reasonable values based on knowledge of the simulation; if the simulation is of fixed photon number $N$, then one should pick $\vec{n}$ such that $\sum_{i=1}^m n_i=N$.
ii) Step 3 can of course be implemented in a variety of manners. If one has a fixed photon number simulation, one can for example, pick two random modes, decrease the photon count in one of the modes by a random amount (if possible), and in the other mode commensurately increase the photon count.
iii) The `thermalization' time $T$ is a user chosen value, to allow the system to (hopefully) reach equilibrium. This will depend on the output distribution, and in principle could be a bottleneck for the procedure (though it depends strongly on the details of the system).
iv) The time complexity for steps 2, 3  is $O(mk)$, for computing the probability, as discussed above. The total cost to output a single measurement result is therefore $O(Tmk)$.

Whilst this method certainly can result in an efficient sampling for practical cases of interest, it has the usual issues and suffers from the same constraints as in most applications of Metropolis sampling. We already briefly discussed the issue of the thermalization time above.
Another issue worth addressing, as also described in Ref.~\cite{Bravyi2019simulationofquantum}, the Markov chain must be irreducible; given two configurations (measurement results) with non-zero probability, there needs to be a path between them, where the path is defined by the type of updates in step 3 above. Whilst in random circuits (such as for Boson sampling) it is likely to have this property (since all measurement results will likely have a non-zero amplitude), in circuits with structure, many amplitudes may be identically 0, leaving the possibility of disconnected regions. Though this can be mitigated to some extent through numerical techniques (dynamically increasing the search area, random restarts etc.), the Metropolis sampling approach may not always be a feasible one.

\subsubsection{Conditional probability sampling}
Here we describe an approach to measurement using conditional probabilities. As before, in principle this could be applied to the coherent state basis also, though the continuous nature adds additional technical details. We will describe the procedure in words.

 Starting with mode 1, output the measurement result $n_1$ with probability $P^{(1)}_{n_1}=\|\langle n_1|\psi\rangle\|^2$ (where $\|\phi\|^2=\| |\phi\rangle\|^2 = \langle \phi|\phi\rangle$). To implement this, one can pick a random number $r\in (0,1)$, and output $n_1$ such that $\sum_{n=0}^{n_1}P^{(1)}_n>r$. Then, with the $m-1$ mode (normalized) state $\ket{\psi_{2,\dots, m}} =\frac{1}{{\|\langle n_1|\psi\rangle\|}} \langle n_1|\psi\rangle$, one applies the same algorithm to the next mode, outputting $n_2$ with probability $\| \braket{n_2}{\psi_{2, \dots m} }\|^2$. Continue until all $m$ modes have been measured. 

The cost of this procedure, to output result $(n_1, \dots, n_m)$  is $O(\sum_{i=1}^m n_i\mathcal{N}_i)$, where $\mathcal{N}_i$ is the cost of computing the norm of the projected state $\braket{n_1, \dots, n_{i}}{\psi}$. 
 As discussed in Sect.~\ref{sect:compute-amps}, the worst case scaling for such a computation is $O(mk^2)$, meaning the cost to output a single measurement sample is $O(Nmk^2)$, where $N=\sum_i n_i$ is the total number of photons. The best or average cases however may be significantly better than this, see Eq.~\eqref{eq:partial-norm}.
 In general though, this is a huge computational overhead, e.g. for single photon inputs, $k=2^n$. (Note, in the case of coherent basis measurements -- heterodyne -- the norm can be computed in a time linear in $k$.)
 
 There are however two possible ways to improve upon this. First, since the overlap between coherent states decreases exponentially in their separation, $|\langle \alpha |\beta\rangle |^2 = e^{-|\alpha - \beta|^2}$, one may be able to pick a representation where all states are approximately orthogonal, in which case one can output a single measurement sample of $N$ photons in time $O(Nk)$ (if the states in the decomposition are approximately orthogonal, the norm can be computed in time $O(k)$). One example as to how this can be achieved follows the Fourier decomposition Eq.~\eqref{eq:fock-fourier}. Consider approximating the Fock state $\ket{n}$. Notice one can increase the error parameter $\epsilon$ (which decreases the overlap between states in the decomposition), and the expense of increasing the rank of the state (e.g. take $N>n$ terms, where the amplitudes $a_i=\delta_{i,n}$ for $i=0, \dots, N-1$). The parameter $\epsilon$ does not, per se, have to be small, since the error goes as $N! \epsilon^{2N+1}/(2N+1)!$, and so can be compensated by increasing $N$. This can allow one to construct a decomposition with a greater number of terms, but where the coherent states in the superposition are approximately orthonormal. 
 Since unitarity preserves the overlap of the initial states in the decomposition, one only needs to guarantee the orthogonality in the initial states construction.

 A second approach is to use a method of estimating the norm, as demonstrated in Ref.~\cite{Bravyi2019simulationofquantum}, for a stabilizer rank decomposition. This relies upon the following proposition:
 \begin{proposition}
     For sufficiently large $L$, the expected overlap of coherent state $\ket{\beta_1, \dots, \beta_m}$, where $|\beta_i|<L$, with an $m$ mode (generally unnormalized) `state' $\ket{\psi}=\sum_{i=1}^k c_i\ket{\alpha_i^{(1)}, \dots, \alpha_i^{(m)}}$ is
     \begin{equation*}
         \mathbb{E}_{\beta_1, \dots, \beta_m:|\beta_i|<L}|\braket{\beta_1, \dots, \beta_m}{\psi}|^2 \approx \frac{1}{L^{2m}} \|\psi\|^2.
     \end{equation*}
\label{prop:coherent-sample-norm}
 \end{proposition}
 \textit{Proof.} We perform the calculation for a single mode (the result generalizes straightforwardly).
 We wish to compute 
 \begin{equation*}
      \mathbb{E}_{\beta : |\beta|<L} |\langle \beta |\psi\rangle|^2 = \frac{1}{\pi L^2}\int_{|\beta|<L} d\beta |\langle \beta|\psi\rangle|^2 
 \end{equation*}
 where $\ket{\psi} = \sum_{i=1}^k c_i |\alpha_i\rangle$. Here the $1/\pi L^2$ is for normalization of the integral over the disk. Since the overlap $|\braket{\beta}{\alpha_i}|^2 = e^{-|\alpha_i - \beta|^2}$ is exponentially small in the distance between the states on the plane, if $L$ is large enough, we can replace the integral by an integral over the entire plane. To be more explicit, we are making the approximation $\langle \beta|\psi\rangle \approx 0$ for $|\beta|>L$, which is in effect a photon number cutoff. In this case, where the integral is now over the entire plane, we see
 \begin{equation*}
 \begin{split}
     \mathbb{E}_{\beta : |\beta|<L} |\langle \beta |\psi\rangle|^2 \approx \frac{1}{\pi L^2}\int d\beta \langle \beta|\psi\rangle \langle \psi |\beta\rangle = \frac{1}{L^2}\mathrm{Tr}[\ketbra{\psi}{\psi}] = \frac{1}{L^2}\|\psi\|^2,
\end{split}
 \end{equation*}
 which is the desired result for $m=1$, where we used the identity $\mathrm{Tr}[A] = \frac{1}{\pi}\int d\beta \langle \beta | A| \beta\rangle$. The approximation becomes arbitrarily accurate for large enough $L$.

 In the $m$ mode case, we just get a product of integrals, and since the trace distributes over tensor products $\mathrm{Tr}[A\otimes B] = \mathrm{Tr}[A]\mathrm{Tr}[B]$, the result is the same, but with normalization $L^{2m}$. \hfill \qedsymbol

 Prop.~\ref{prop:coherent-sample-norm} implies yet another method for computing the norm linearly in the rank, in principle. In particular, since each overlap squared $|\braket{\beta_1, \dots, \beta_m}{\psi}|^2$ costs $O(mk)$ steps to compute (as previously discussed), the cost for estimating the norm is $O(Tmk)$, where $T$ is the number of random samples required to estimate the norm to desired accuracy. Combining this with the above description for conditional sampling, the cost for outputting a single measurement sample of $N$ photons is $O(NTm k)$.

\begin{figure}
    \centering
    \includegraphics[width=0.9\columnwidth]{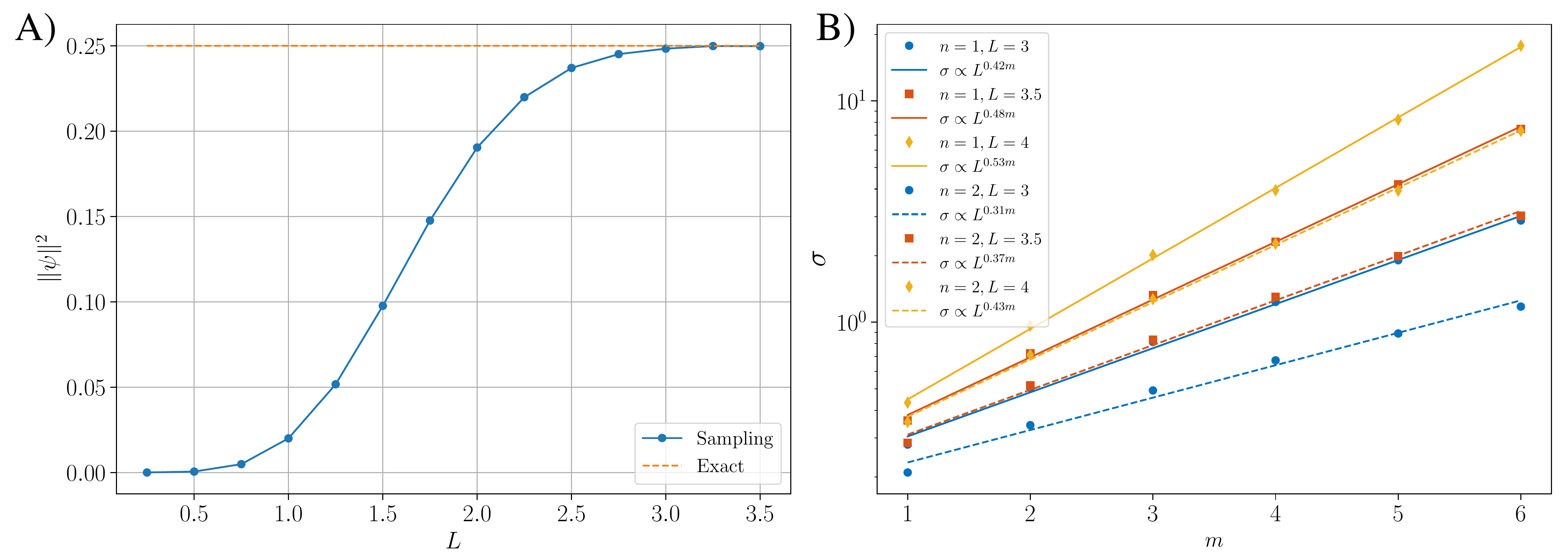}
    \caption{\textbf{Estimating the norm via Prop.~\ref{prop:coherent-sample-norm}}. A) We estimate the norm squared for a rank three unnormalized state $\frac{1}{2} \ket{2}$ (using Cor.~\ref{cor:fock} with $\epsilon=0.35$), taking 750000 samples per data point. B) Numerical computation of the standard deviation from 750000 samples, for states of the form $\frac{1}{2}\ket{n}^{\otimes m}$, for $n=1,2$ in the legend (i.e. states with norm squared 1/4). We use Cor.~\ref{cor:fock} to write $\ket{n}^{\otimes m}$ as a superposition of coherent states with rank $(n+1)^m$, where we use error parameter $\epsilon=0.35$. We also vary the sampling radius $L$. The straight lines are least squares fits to the data the form $\sigma = A L^{c m}$, with $c$ shown in the legend.}
    \label{fig:norm-est}
\end{figure}

 Of course, there are some potential practical issues to address. First, the choice of $L$ is critical; if it is too small, the estimate will not converge to the correct value, and if it is too large, the number of samples $T$ required will also be large. To pick $L$, one could start with a small value, and increase it until convergence, but of course this costs additional resources.
In Fig.~\ref{fig:norm-est} A) we show an example of estimating the norm for a single mode state using Prop.~\ref{prop:coherent-sample-norm}. We see in this case, $L\approx 3$ suffices to get within 1\% accuracy of the true value.

Another obvious issue is the number of samples required. First, let's define the random variable $X = L^{2m} |\langle \vec{\beta} |\psi\rangle|^2 $, where $\ket{\vec{\beta}} = \ket{\beta_1, \dots, \beta_m}$. By Prop.~\ref{prop:coherent-sample-norm} we know $\langle X\rangle_\beta \rightarrow \|\psi\|^2$, for a large enough sampling radius $L$. The variance, for sufficiently large $L$, is given by
\begin{equation}
    \sigma^2 = \langle X^2\rangle -\langle X \rangle^2 \approx \frac{L^{2m}}{\pi^{m}} \int |\langle \vec{\beta}|\psi\rangle|^4 d\vec{\beta} - \|\psi\|^4.
\end{equation}
Here we use the integral is normalized by the area $\pi L^2$ for each mode. Unfortunately, this will nominally scale exponentially in $m$. For example, if $\ket{\psi}$ is a tensor product of coherent states, the first term will be proportional to $(L^2/2)^m$, as each of the $m$ integrals is just a Gaussian, which unless $L \le \sqrt{2}$ will result in an exponential sampling overhead. 
We also demonstrate this numerically in Fig.~\ref{fig:norm-est} B) for Fock basis states, in these cases with worst case $\sigma \sim L^{0.5 m}$. Such scaling would result in sampling complexity $T = O(L^m)$ (since the standard error scales as $1/\sqrt{T}$).
Nevertheless, the precise scaling will depend on the details of the system being simulated, and depending on the choice of $L$ for convergence to a desired accuracy (which in principle can itself depend on $m$), more efficient sampling protocols may still be feasible (e.g., if one is able to scale $L=\ell^{1/m}$).

It is interesting to note that the finite dimensional (qubit) analogue of this does not suffer such a sampling overhead, since the classical states -- stabilizer states -- form a 2-design and thus sampling fluctuations (standard deviation) are suppressed by the Hilbert space dimension \cite{bravyiImprovedClassicalSimulation2016, Bravyi2019simulationofquantum, lowLargeDeviationBounds2009}. In contrast, it is known that any set of CV states can at best form a 1-design (indeed, coherent states form a 1-design, as they resolve the identity) \cite{no-2-deisgn-cv, albert-cv-t-design}.

\subsubsection{Summary of measurement results}
 We sum up the results of this section in the following table for Fock basis measurements, but direct the reader to the relevant sections above for a more nuanced picture. Note, the first row does not correspond to outputting statistics according to the measurement distribution, it is the cost for simply computing a single probability (the other rows do however correspond to outputting results according to the measurement distribution).

\begin{center}
\begin{tabular}{ |p{6cm}||p{3cm}|p{4.5cm}|  }
 \hline
 \multicolumn{3}{|c|}{Measurement procedure cost for $m$-mode rank-$k$ state} \\
 \hline
 Method & Time complexity & Notes\\
 \hline
 Single measurement probability & $mk$ & \\
 Markov chain sampling   & $Tmk$    & Thermalization time $T$\\
 Conditional probability, exact norm &   $Nm k^2$  & Number photons $N$  \\
  Conditional probability, approx. ortho. & $Nk$ & \\
 Conditional probability, Prop.~\ref{prop:coherent-sample-norm}    & $NTmk $ & Number samples $T=T(L)$\\
 \hline
\end{tabular}
\end{center}

\subsection{Comment on errors}
Since the framework described allows for a certain degree of imprecision in a decomposition (e.g., Fock states are only approximately represented, as in Th.~\ref{th:general-fock}, albeit to arbitrary precision), a natural question is the propagation of errors. 
If the initial approximate state $\ket{\tilde{\psi}}$ has target fidelity $|\braket{\tilde{\psi}}{\psi}|^2 = 1-\epsilon$ with the ideal state $\ket{\psi}$, under free unitary operations the overlap remains unchanged. 
For an initial product state $\ket{\tilde{\psi}}^{\otimes n}$, the fidelity is $(1-\epsilon)^n$. In order to achieve a total fidelity $1 - \delta$, one can take $\epsilon = 1 - (1-\delta)^{1/n} \approx \delta/n$ for small $\delta$ (i.e. it only requires linearly decreasing precision per state).
Of course, if one implements resourceful operations inexactly (as in Cor.~\ref{cor:creation_cost}), errors can be introduced during the simulation, however, these can be controlled easily by picking the error parameters appropriately.

\begin{figure}
    \centering
    \includegraphics[width=0.7\columnwidth]{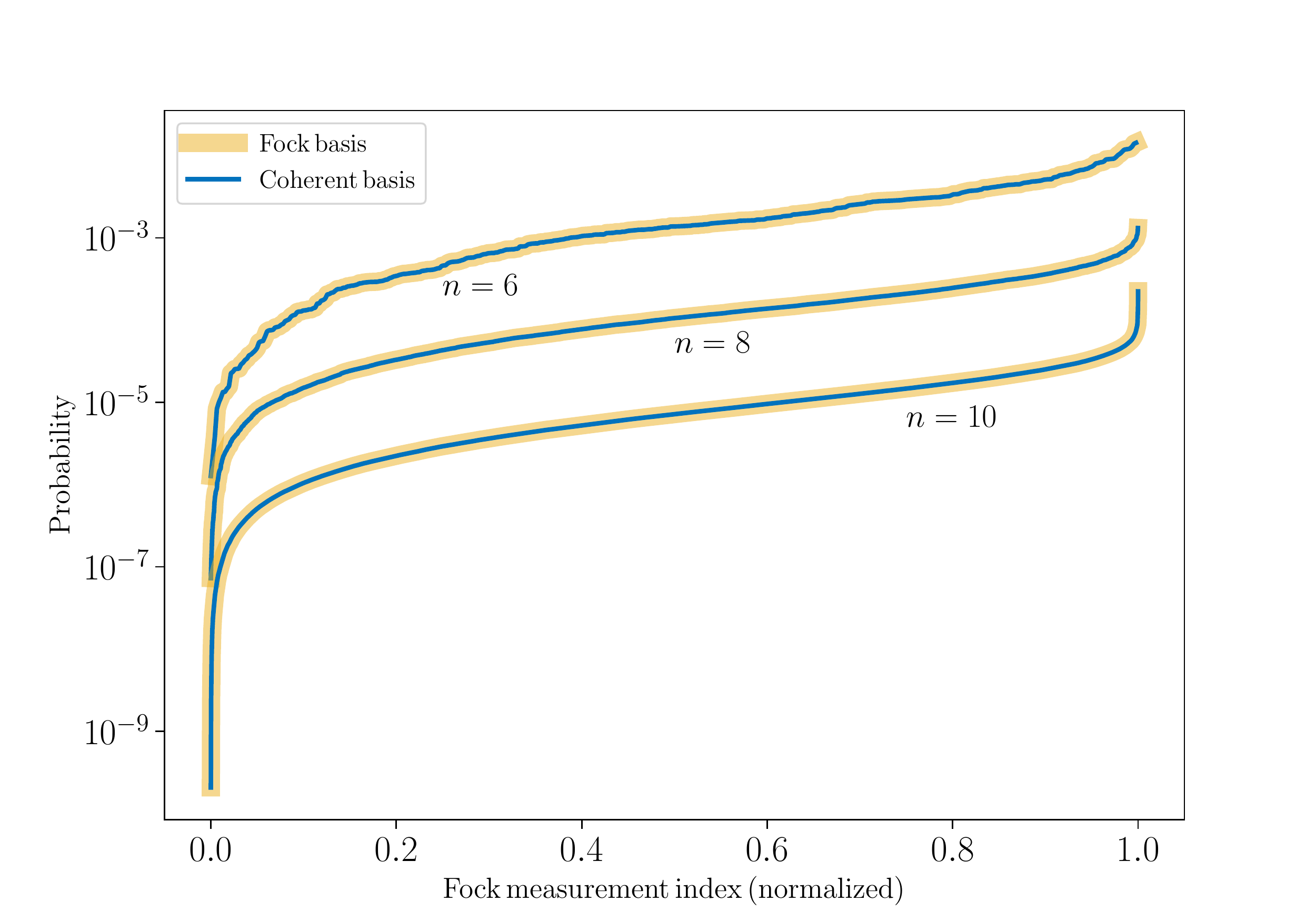}
    \caption{\textbf{Boson sampling simulation}. Starting with initial state $\ket{1}^{\otimes n}$, where $n=6,8,10$, we simulate the output from a Haar random unitary transfer matrix. In the Fock basis, this results in storing all 462, 6435, 92378 complex amplitudes respectively (i.e., the dimension of the Hilbert space for the $n$ photons in $n$ modes). The coherent state decomposition uses Cor.~\ref{cor:fock} to represent $\ket{1}$ as an odd cat state, with error parameter $\epsilon=0.2$, which has fidelity $|\langle 1|\tilde{1}\rangle|^2 > 0.999$. This results in $(n+1)2^n= 448, 2304, 11264$ complex numbers stored in memory respectively. We compute the Fock probabilities via Eq.~\eqref{eq:fock_coherent_amps}, which are enumerated on the x-axis, sorted by the probability.}
    \label{fig:boson-sample}
\end{figure}

In Fig.~\ref{fig:boson-sample} we show the output measurement probabilities from a Boson sampling like simulation, for sizes up to 10 photons in 10 modes, comparing the results from an exact simulation in the Fock basis (as outlined in Sect.~\ref{sect:fock-sim}), to simulation using the coherent rank decomposition. We see the results match very well, for all output states, where the initial fidelity is set to over 99.9\% for each single photon state.

Lastly we discuss the issue of numerical stability. We take the canonical example of starting in a state with $n$ single photons $|1\rangle^{\otimes n}$ (though any other $n$ photon configuration would work equally well). In the coherent framework this is represented as a product of odd cat states, which for small error $\epsilon$ can be written
\begin{equation}
    |1\rangle^{\otimes n} \approx \frac{1}{(2\epsilon)^n} \sum_{i=1}^{2^n} s_i |\epsilon \vec{b}_i\rangle,
    \label{eq:state_sum}
\end{equation}
where $\vec{b_i}$ is a binary vector of $\pm 1$ entries, and $s_i =\prod_j b_{i,j} \in \{\pm 1\}$, where $b_{i,j} = (\vec{b}_i)_j$ is the $j$'th component of vector $\vec{b}_i$. That is, the coherent states in the superpositon are $\vec{\alpha}_{i} = \epsilon \vec{b}_i = (\epsilon b_{i,1}, \dots, \epsilon b_{i,n})$.
It is evident for small enough $\epsilon$ and/or large enough $n$, the pre-factor could be numerically unstable, and lead to inaccurate computations. We can observe however that this need not be explicitly stored, and used mostly as a `bookkeeping' device. 
Indeed, consider updating such a state by a linear optical unitary $\hat{U}$, with transfer matrix $\hat{u}$. We can notice this can be performed agnostic of $\epsilon$:
\begin{equation*}
    \hat{U}|1\rangle^{\otimes n}  \approx \frac{1}{(2\epsilon)^n} \sum_{i=1}^{2^n} s_i |\epsilon \hat{u}\vec{b}_i\rangle = \frac{1}{(2\epsilon)^n} \sum_{i=1}^{2^n} s_i |\epsilon \vec{b}_i'\rangle.
\end{equation*}
Here, by slight abuse of notation, $\vec{b}_i' = \hat{u}\vec{b}_i$ is the unitarily evolved vector (e.g., see Eq.~\eqref{eq:general-lo-update}), which is no longer a `binary' vector in general. At this step, only the vectors $\vec{b}_i'$ (and corresponding phase factors) need to be stored explicitly. Moreover we see there is no issue of compounding errors when additional unitaries are applied (the vectors $\vec{b}_i'$ can just be updated, which contain no reference to $\epsilon$).

When computing a transition amplitude (again with a slight abuse of notation) $\langle \vec{n}|\hat{U}|1\rangle^{\otimes n}$, we see the $\epsilon$ dependence can be almost dropped entirely:
\begin{equation*}
    \langle \vec{n}|\hat{U}|1\rangle^{\otimes n}  \approx  \frac{1}{(2\epsilon)^n} \sum_{i=1}^{2^n} s_i \langle \vec{n} | \epsilon  \vec{b}'_i\rangle = \frac{1}{2^n}\sum_{i=1}^{2^n} s_i \prod_{j} e^{-\epsilon^2 |b_{i,j}'|^2/2}  \frac{(b_{i,j}')^{n_j}}{\sqrt{n_j!}}.
\end{equation*}
In the last step we used that $\ket{\vec{n}}$ is an $n$ photon state $n=\sum_j n_j$ and so the overlaps with the coherent states cancel out the $\epsilon^n = \prod_j \epsilon^{n_j}$ contribution. In fact in this instance, one could formally set $\epsilon = 0$ in the last step to recover the \textit{exact} result.

In more general scenarios, this kind of `bookkeeping' may not be desired/feasible in which case there could, in principle, be numerical precision issues at large sizes. This of course depends strongly on the desired accuracy.  
In Fig.~\ref{fig:boson-sample}, we see $\epsilon = 0.2$ gives high accuracy (above $99\%$) to all output probabilities. In Eq.~\eqref{eq:state_sum} this yields a prefactor of $2.5^n$, which would likely pose no numerical problems, at the accessible system sizes (i.e. those that fit in memory). Overall, the value $\epsilon$ should be chosen carefully to avoid numerical issues at large sizes. Fortunately, the  fidelity of the approximate state converges quickly as $O(\epsilon^{2(n+1)}/(n+1)!)$ for an $n$-photon single mode state, as shown in Th.~\ref{th:general-fock}. Unless very large sizes, and extremely high precision is required, one can likely avoid any such issues.

\section{Resource theory of coherent state rank \label{sect:resource_th}}
The framework introduced in this work can be phrased as a quantum resource theory \cite{chitambarQuantumResourceTheories2018}. As a brief reminder, a resource theory is a triple, \((\mathcal{F}, \mathcal{O}, \mathcal{R})\), where: \(\mathcal{F}\) is the set of `free states,' a subset of quantum states \(\rho \in \mathcal{S}(\mathcal{H})\) that are devoid of the resource of interest\footnote{Here \(\mathcal{S}(\mathcal{H})\) is the state space, i.e., the set of all density matrices}. \(\mathcal{O}\) is the set of `free operations,' a subset of quantum channels, \(\mathcal{E}: \mathcal{S}(\mathcal{H}) \rightarrow \mathcal{S}(\mathcal{H})\) that do not generate any resource when applied to states in \(\mathcal{F}\). And \(\mathcal{R}:\mathcal{S}(\mathcal{H}) \rightarrow \mathbb{R}_{\ge 0 }\) is a functional quantifying the amount of resource in quantum states, with certain properties such as, (i) faithfulness, i.e., vanishing on the resource-free states only: \(\mathcal{R}(\rho) = 0 \iff \rho \in \mathcal{F}\) and (ii) monotonicity under free operations, i.e., free operations can at most \textit{consume} the resource but never increase it: \(\mathcal{R}(\mathcal{E}(\rho)) \leq \mathcal{R}(\rho) ~~\forall \mathcal{E} \in \mathcal{O}\).

In our formulation, free states are multi-mode coherent states of the form $\ket{\alpha_1, \dots, \alpha_m}$, where $\ket{\alpha_i}$ is a coherent state, Eq.~\eqref{eq:coherent-state}. Free operations are those described in Sect.~\ref{sect:free-ops}, which map (multi-mode) coherent states to coherent states (up to the amplitude), such as beamsplitters, phase shifts, displacement operators. Then, akin to the definition of ``magic" or ``non-stabilizerness" in stabilizer rank simulations \cite{Veitch_2014-resource-stabilizer, stab-renyi,quantifying-speedup}, a resource measure for an arbitrary $m$ mode state $\ket{\psi}$ is the logarithm of the rank of its minimal decomposition into free states. That is\footnote{Logarithm base is unimportant, but we will use base 2 here} $\mathcal{R}(\ket{\psi})=\log k$, where $k$ is the smallest positive integer such that $\ket{\psi}=\sum_{i=1}^k c_i \ket{\alpha_i^{(1)}, \dots, \alpha_i^{(m)}}$.
Properties i) and ii) are immediately satisfied by this definition. Additionally, it is easy to check this satisfies subadditivity, $\mathcal{R}(\ket{\psi}\otimes \ket{\phi}) \le \mathcal{R}(\ket{\psi}) + \mathcal{R}(\ket{\phi})$, and in particular $\mathcal{R}(\ket{\psi}\otimes \ket{\phi}) = \mathcal{R}(\ket{\psi})$ for $\ket{\phi}\in \mathcal{F}$.

We emphasize that our focus in this work is primarily on resource quantification and not on state conversion. This allows us to restrict to pure states and free unitaries, while acknowledging the fact that free unitaries cannot consume any resource, see App.~\ref{sec:free-unitaries}. A similar notion related to our coherent rank resource framework is the so called `degree of non-classicality', which has previously been studied in e.g., Refs.~\cite{gehrke-quantification, vogel_unified-quantification}.

Moreover, the coherent state rank is closely related to the notion of \textit{coherence rank} in the resource theory of coherence \cite{streltsovColloquiumQuantumCoherence2017}. For example, if the coherent states in the expansion of a quantum state are nearly orthogonal then the two are equivalent. However, generically, coherent states are \textit{not} orthogonal and therefore a direct connection to resource theory of coherence is slightly more involved. The quantification of coherence in infinite-dimensional systems is not a new problem. Two approaches are quite common here: the first assumes Fock states as the ``incoherent basis"\footnote{Slightly unfortunate jargon here but we hope the reader can distinguish the resource theoretic incoherent/coherent states from the \textit{optical} coherent states.} and quantifies the coherence of coherent states with respect to this, e.g., Refs. \cite{PhysRevA.93.012334,PhysRevA.93.032111}. However, this has the disadvantage of being at odds with the traditional notion of non-classicality in quantum optics, where coherent states are the most classical, while Fock states are admittedly non-classical \cite{gerry_knight_2004}. The second approach defines coherent states as the preferred incoherent basis. This includes the resource theory of superposition \cite{theurerResourceTheorySuperposition2017} and subsequent works \cite{tanQuantifyingCoherenceCoherent2017}, including an operational framework \cite{yadin_operational_2018}. Our resource theory framework falls under the so-called passive linear optics framework and the operational characterizations introduced in the references above are broadly applicable to our work.

In the present work, we consider the coherent state rank as the quantity of interest, since this directly determines the classical simulation complexity.
We add one comment here however, that since the decomposition of an arbitrary state $\ket{\psi}$ to a coherent state basis is typically not exact for finite $k$ (e.g. see Th.~\ref{th:general-fock} and Eq.~\eqref{eq:coherent-integral}), we will allow the above resource theory to hold approximately, in the sense of Def.~\ref{def:approx-coherent-rank}.

\begin{figure}
    \centering
    \includegraphics[width=0.6\columnwidth]{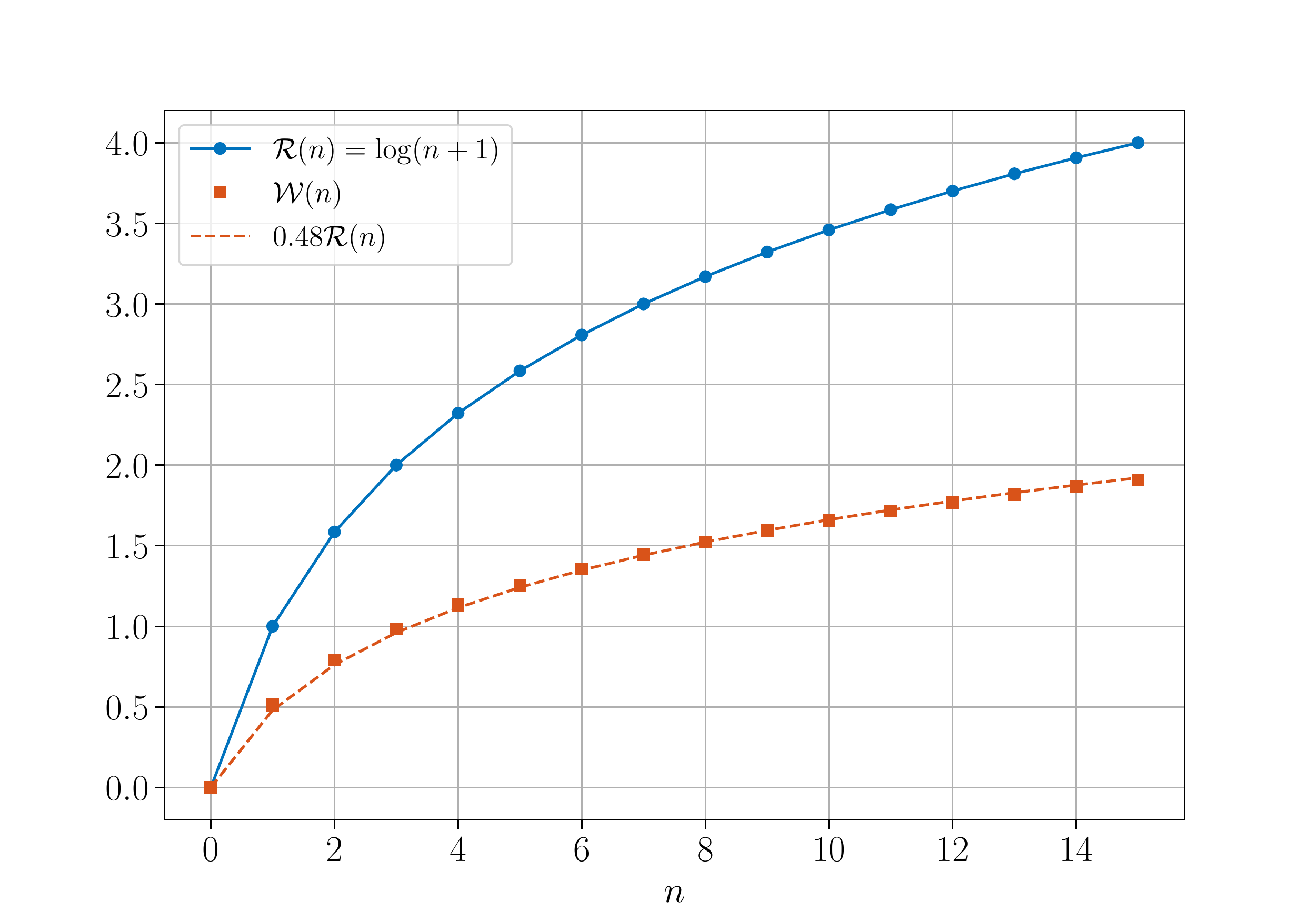}
    \caption{\textbf{Resource monotones for Fock states}. We plot the measures of resource for the Fock states $\ket{n}$, using the coherent state rank definition, $\mathcal{R}$, and the Wigner negativity resource (with logarithm base 2). We also perform a curve fit (dash line), which shows the Wigner resource is approximated by $\mathcal{W}(n)\approx 0.48 \mathcal{R}(n)$. In App.~\ref{sect:fock-wigner-neg} we demonstrate by exact calculation that indeed the result is only approximate, and likely the similar growth is due to the fact $W_n(\alpha)=W_n(|\alpha|)$ has $n$ roots in $|\alpha|$.}
    \label{fig:fock-resource}
\end{figure}

We can contrast the coherent state rank resource theory to another related resource theory, namely one using the monotone of the Wigner logarithmic negativity \cite{wigner-neg-resource-th}, defined as $\mathcal{W}(\ket{\psi}) = \log \int d\vec{\alpha} |W_{\ket{\psi}} (\vec{\alpha})|$, where $\vec{\alpha} =  (\alpha_1, \dots, \alpha_m)$ for an $m$-mode pure state $\ket{\psi}$.
In this framework, Gaussian states (including coherent states) are free, as the Wigner function is everywhere non-negative.
First note that this implies  $\mathcal{W}(\ket{\psi})=0=\mathcal{R}(\ket{\psi})$ for $\ket{\psi}\in \mathcal{F}$.
Moreover, in Fig.~\ref{fig:fock-resource} we plot both resource measures for Fock states, which shows the two definitions are consistent (in fact, are related approximately by a constant factor). Here we use the results of Th.~\ref{th:general-fock} that shows one can approximate $\ket{n}$ using $n+1$ coherent states\footnote{This leaves open the possibility of more efficient coherent state decompositions of Fock states, so $\log(n+1)$ is technically an upper bound for $\mathcal{R}(\ket{n})$}.
However, it is also clear the two definitions will not generally align with one another, in the sense one can  find states $\ket{\psi}, \ket{\phi}$ such that $\mathcal{W}(\ket{\psi})<\mathcal{W}(\ket{\phi})$ but $\mathcal{R}(\ket{\psi})>\mathcal{R}(\ket{\phi})$, or vice versa (this is easy to show in the case we consider exact decompositions\footnote{E.g., take (normalized) versions of $\ket{\psi}\propto \ket{\epsilon} + \ket{2\epsilon} + \ket{3\epsilon}, \ket{\phi}\propto \ket{\epsilon} - \ket{-\epsilon}$. Then, for certain small enough values of $\epsilon$, in the Wigner setting $\ket{\phi}$ will have a larger resource, since $\ket{\psi}$ ($\ket{\phi}$) is close to $\ket{0}$ ($\ket{1}$), whereas the coherent rank decomposition of $\ket{\psi}$ is larger.}, but harder to see when we allow approximate states, due to the difficulty of proving the optimality of the approximate rank $k$ of a particular state decomposition, though we expect the statement to remain true).

Similarly, whilst the coherent rank decomposition is related to the stellar rank via Prop.~\ref{prop:stellar-coherent-ranks}, we wish to make it clear these are also distinct theories. 
In the context of a resource theory, the stellar rank $r$ is a genuine resource monotone where Gaussian states and operations are free ($r=0$), and simulation complexity scales as $O(2^r)$ (Sect.~\ref{sect:stellar-sim}). 
By Prop.~\ref{prop:stellar-coherent-ranks}, one can see the approximate coherent rank resource measure
\begin{equation*}
    \mathcal{R}_\psi \le \log [(r_\psi+1)] + \log [k_\psi],
\end{equation*}
where $k_\psi$ is the coherent rank of the Gaussian state in the stellar decomposition of $\ket{\psi}$, Eq.~\eqref{eq:psi-stellar}.
Two comments are in order. First, notice that the approximate coherent rank \(\mathcal{R}_\psi \) is upper bounded by two different types of \textit{non-classicality}. The stellar rank term, \(\log [(r_\psi+1)]\) represents the \textit{non-Gaussian} operational cost in terms of single-photon additions required to prepare the quantum state \cite{stellar-rep-non-gaussian}, where this term is vanishing if and only if the stellar rank is zero. On the other hand, the  \(\log [k_\psi]\) term denotes the non-classicality in terms of the Glauber-Sudarshan P function (see e.g., Ref.~\cite{tanQuantifyingCoherenceCoherent2017}), related to the amount of squeezing\footnote{Recall, a single mode Gaussian state can be written $\hat{S}(\zeta)\ket{\alpha}$, for some $\alpha, \zeta \in \mathbb{C}$, meaning a Gaussian state with non-trivial coherent state rank ($k>1$) is entirely due to non-zero squeezing.}, which is vanishing if and only if \(k_\psi = 1\), when the squeezing is $\zeta=0$. Second, as a function of the stellar rank, the \(\mathcal{R}_\psi\) can grow \textit{at most} logarithmically. However, the overall scaling of \(\mathcal{R}_\psi \) may be \textit{dominated} by the amount of squeezing instead, since, \(k_\psi\) can itself be arbitrarily large, for $|\zeta| \rightarrow \infty$ (see also the discussion after Prop. \ref{prop:stellar-coherent-ranks}).

In terms of simulation complexity, the classical requirements for Fock input state $\ket{1}^{\otimes n}\ket{0}^{\otimes (m-n)}$ under linear optics and with coherent basis measurements, scales as $O(2^n)$ in both the stellar and coherent rank frameworks, though the former scales as $O(4^n)$ for Fock basis measurements (and similarly for linear combinations of Gaussians \ref{sect:linear-gaussians}). Indeed, it is easy to find differences.
As another example, the state $\ket{n}\ket{0}^{m-1}$ can be simulated in a time linear in $n$ within the coherent state framework, whereas this is a state with stellar rank $n$ and simulation complexity therefore scaling as $2^n$ in the stellar rank paradigm (Sect.~\ref{sect:stellar-sim}). Another key example is that the cat state $\ket{ix}-\ket{-ix}$ has coherent rank exactly 2, whereas the stellar rank is infinite \cite{stellar-rep-non-gaussian}. On the other hand, Gaussian Boson sampling is typically much more efficient in the stellar framework (though as discussed below, for small enough squeezing or low enough target fidelity, the two can be equivalent, but when the squeezing is large, the stellar formalism will be most efficient).

Nevertheless, in all three cases, Wigner logarithmic negativity, coherent state rank, and the stellar rank, the resource measure, $\mathcal{W}$, $\mathcal{R}$, and $r$ relates to the simulation cost. For coherent state rank and stellar rank, the cost is exponential, scaling as $2^{\mathcal{R}, r}$, for free operations. For the Wigner negativity, as discussed in App.~\ref{sect:sign-problem}, the simulation complexity scales as $4^{\mathcal{W}}$, although this has additional difficulties/errors as it is a sampling approach.

\section{Discussion \label{sect:discussion}}
We have developed a paradigm for the simulation of quantum optics, by decomposition to coherent states, in a manner akin to stabilizer state decompositions in finite dimensional systems. In this framework, linear optical operations are free, as are computing amplitudes in the Fock or coherent state bases. For Boson sampling type simulations, this can lead to exponential improvements over naive Fock space simulation, with a simulation cost comparable to methods relying upon the permanent.

Below we summarize some of the complexity results for a few examples using the finite coherent state rank decomposition discussed in this work. Starting with an initial state of $n$ photons, we document the cost for storing this in memory (space), and the update time for a $m-$mode linear optical (LO) operation (e.g., random Boson sampling circuit). Here the measurement column is the time cost to compute the probability of a particular measurement outcome, in the Fock or coherent state basis. All results in the table should be interpreted in the `big O' style, where e.g. lower order terms and constant factors related to implementation details are hidden.
In the fourth row down, $\ket{0.882}$ refers to a squeezed vacuum state (e.g., Gaussian Boson sampling, GBS) with squeezing parameter $\zeta=0.882$, corresponding to mean photon number $\langle \hat{n}\rangle \approx 1$. We find such a state can be approximated to fidelity above 0.99 using 8 coherent states, which is used here (for lower/higher target fidelity, fewer/more states are needed in general). Also note, for larger (smaller) squeezing, typically more (fewer) coherent states will be required in the decomposition.
\begin{center}
\begin{tabular}{ |p{3.75cm}|p{1.25cm}|p{1.5cm}|p{2.25cm}|p{4cm}|  }
 \hline
 \multicolumn{5}{|c|}{Complexity overview for $n$ photon simulations} \\
 \hline
 Initial state & Space & LO Time & Measurement & Notes  \\
 \hline
  $\ket{\alpha=1}^{\otimes n}\ket{0}^{\otimes (m-n)}$ &   $m$ & $m^2$ & $m$  &  Coherent state $\ket{\alpha}$ \\
 $\ket{1}^{\otimes n}|0\rangle^{\otimes {(m-n)}}$   & $m2^n$  & $m^2 2^n$ & $m2^n$  &  Boson sampling\\
 $\ket{n}\ket{0}^{\otimes (m-1)}$ &   $mn$ & $m^2 n$ & $mn$  & Single-mode Fock state \\
 $\ket{\zeta=0.882}^{\otimes {n}}\ket{0}^{\otimes {(m-n)}}$ & $m 8^n$ & $m^2 8^n$  & $m8^n$  & Squeezed state $\ket{\zeta}$ to 99\% fidelity (GBS)\\
 \hline
    $\sum_{i=1}^k c_i \ket{\alpha_i^{(1)}, \dots, \alpha_i^{(m)}}$ &   $mk$ & $m^2 k$ & $mk$  &  General rank $k$ state \\
 \hline
\end{tabular}
\end{center}
In all cases in the above table, the simulation requirements here can be much less than the equivalent full Fock space simulation (Sect.~\ref{sect:fock-sim}), which scales as the full Fock dimension in the worst case (such as in Boson sampling). Notably, the simulation cost here is only quadratic in the number of modes, even if all modes of an $n$ photon system are populated. This allows us to represent such systems significantly more efficiently, based on the decomposition used. 

The method introduced is trivially parallelizable, since different compute nodes can store a subset of the terms in a coherent state decomposition, and update them independently under linear optical operations. Moreover, this method potentially allows one to construct a trade off between simulation complexity and accuracy, e.g., by reducing the number of terms in a coherent state decomposition, at the expense of fidelity. In the above table for example, we use 8 coherent states to approximate a particular squeezed state to fidelity 0.99, however fewer terms are required in general for a lower target fidelity; for this case, $\zeta=0.882$, only 2 coherent states are required to achieve fidelity above 0.9, yielding a cost to simulate GBS scaling as $2^n$. This allows a heuristic approach to simulating Boson sampling up to a target fidelity.

Additionally this construction allows for a tensor network like approach, potentially resulting in memory (and time) saving, in circuits with structure \cite{vidal-efficient-2003, markov_quantum_2018}. For example, in systems that have only a few entangling operations across a particular partition, we can split the state as a sum of tensor product states, across this partition. Starting with a product state, one can always write $\ket{\psi} = \ket{\psi_L}\otimes \ket{\psi_R}$. Let's assume for simplicity equal partitions, so each partition size (number of modes) is half of the full one (i.e., dimension $\sqrt{d}$ instead of $d$). Instead of storing one state over the entire Hilbert space $O(d)$, we use the structure to store two states, but each only over their respective half partition, $O(\sqrt{d})$. In the coherent rank picture for say Boson sampling with $n$ photons, the cost to store this in memory is only $O(2^{n/2})$, instead of $O(2^n)$ for a state over the full Hilbert space.
Then, any operation acting only on the left or right half can be applied trivially. For entangling operations (e.g., two-mode operations acting across the partition), if they can be decomposed as $\hat{U} = \sum_{i=1}^kc_i \hat{u}_i \otimes \hat{v}_{i}$, where $\hat{u}_i, \hat{v}_i$ are free (e.g., displacements) acting on each partition, then the state becomes a superposition of size $k$, but preserving the partition structure in each term.
In the coherent rank decomposition picture, writing $\ket{\psi_L} = \sum_{m=1}^{k_L} a_m \ket{\vec{\alpha}_m}$, $\ket{\psi_R} = \sum_{n=1}^{k_R} b_n \ket{\vec{\beta}_n}$, we initially just store $k_L + k_R$ states in memory (here $\ket{\vec{\alpha}_m}$ ($\ket{\vec{\beta}_n}$) is only defined over the modes in the left (right) partition). Upon application of $\hat{U}$ across the partition, we get
\begin{equation}
\begin{split}
    & \hat{U}\ket{\psi} = \sum_{i=1}^k c_i \left( \sum_{m=1}^{k_L} a_{m,i} \ket{\vec{\alpha}_{m,i}} \right) \otimes \left(\sum_{n=1}^{k_R} b_{n,i}   \ket{\vec{\beta}_{n,i}}\right)=:\sum_{i=1}^k c_i \ket{\psi_{L,i}}\otimes \ket{\psi_{R,i}}, \\ 
    & \hat{u}_i \ket{\vec{\alpha}_m} = \frac{a_{m,i}}{a_m}\ket{\vec{\alpha}_{m,i}},\; \hat{v}_i \ket{\vec{\beta}_n} = \frac{b_{n,i}}{b_n} \ket{\vec{\beta}_{n,i}},
    \label{eq:state-split-evolve}
\end{split}
\end{equation}
and the cost to store this in memory is $O(k(k_L + k_R))$ from storing each $\ket{\vec{\alpha}_{m,i}}, \ket{\vec{\beta}_{n,i}}$ (and amplitudes), which could be significantly less than $O(k_Lk_R)$ for storing the complete state. Of course, finding a decomposition for $\hat{U}$ could be non-trivial itself, or result in a huge number of terms ($k$).
Further notice that the cost will be exponential in the number of such operations, e.g. after $T$ of them with space complexity scaling as $O(k^T (k_L + k_R))$. In general therefore this is only useful if there are few of such entangling operations across the partition.
Finally we comment that the $\hat{u}_i, \hat{v}_i$ do not strictly speaking have to be free operations, but could also increase the rank in the state of the respective partitions. For example, a beamsplitter (Eq.~\eqref{eq:beamsplitter}) with a small transmission to the other mode (small $\theta$) can be expanded to first order $\hat{B}(\theta, \phi) \approx 1 + \frac{\theta}{2}(\hat{a}^\dag \hat{b} e^{i\phi} - \hat{a}\hat{b}^\dag e^{-i\phi})$. Notice, this has the desired structure, but the creation operators are not free, and instead double the coherent rank, by Cor.~\ref{cor:creation_cost} (the annihilation operators are free). The effect this has would be to increase the $k_L, k_R$ for each term in Eq.~\eqref{eq:state-split-evolve}. For this example, the resulting number of states to keep in memory for each of the three terms would be respectively $k_L + k_R, 2k_L + k_R, k_L + 2k_R$, or $4(k_L + k_R)$ in total. If we kept up to order $p$ in the expansion (i.e., up to terms of the form $\hat{a}^{\dag p}\hat{b}^p$, etc.), the total number of states to store in memory would be\footnote{This calculation follows using Th.~\ref{th:creation_annihilation_cost}, which shows application of the $n$'th order term, $(\hat{a}^\dag \hat{b}e^{i\phi} - \hat{a}\hat{b}^\dag e^{-i\phi})^n$, generates a total of $2^{n-1}(2+n)$ states in each partition (by counting the contribution of each of the $2^n$ terms). The final result is obtained by summing this from 0 to $p$.} $2^p (p+1)(k_L + k_R)$. Whether or not this is useful of course depends strongly on the circuit at hand. We can see if there are only a few beamsplitters applied across the partition of choice, and $p$ ($\theta$) is sufficiently small, a memory saving can be achieved.

\section{Acknowledgements}

We thank Ulysse Chabaud for providing additional references and detailed comments on the first version of this manuscript.

We are grateful to support from DARPA under IAA 8839 Annex 129. J.M. is thankful for support from NASA Academic Mission Services, Contract No. NNA16BD14C. N.A. is a KBR employee working under the Prime Contract No. 80ARC020D0010 with the NASA Ames Research Center. The United States Government retains, and by accepting the article for publication, the publisher acknowledges that the United States Government retains, a nonexclusive, paid-up, irrevocable, worldwide license to publish or reproduce the published form of this work, or allow others to do so, for United States Government purposes.

\bibliography{refs.bib}

\appendix

\section{Sign problem for QPD sampling \label{sect:sign-problem}}

As briefly mentioned in Sects.~\ref{sect:efficient-sim}, \ref{sect:resource_th}, it is possible to efficiently sample from a POVM when the Wigner function of the state and the POVM elements are guaranteed to be non-negative \cite{positive_wigner_sim-Eisert, Veitch_2013}. In fact, the situation can be generalized to quasi-probability-distributions (QPDs) beyond the Wigner function, though the result is similar \cite{QPD-sim_Caves}. We will outline the case for Wigner simulation here for ease of exposition. After this, we will show an instance of the so called sign problem, which arises when there are negative values present in the Wigner function.

We will follow the protocol of Ref.~\cite{positive_wigner_sim-Eisert}, though we will work in the single mode case for simplicity (also note that we differ by a factor of $\sqrt{2}$ in the definition of the phase space, leading to a different normalization condition, but this is totally unimportant). 
We remind the reader of the following identities,
\begin{equation}
   \mathrm{Tr}[A]=\int W_A(\alpha) d\alpha;\; \mathrm{Tr}[AB] = \pi \int W_A(\alpha)W_B(\alpha)d\alpha,
   \label{eq:wigner-properties}
\end{equation}
where all integrals are over the plane: $\int d\alpha = \int_{-\infty}^\infty \int_{-\infty}^\infty d\mathrm{Re}(\alpha)d\mathrm{Im}(\alpha)$.

First we define a POVM $\{M_k\}_{k=0}^K$, where by definition $\sum_{k=0}^K M_k = \mathbb{I}$. Then it is clear measurement result $k$ occurs with probability $P_k = \pi \int W_\rho(\alpha) W_k(\alpha)d\alpha$, where $\rho$ is the state we are measuring, and $W_k$ the Wigner function for $M_k$. 
By definition, we also have $\sum_{k=0}^K W_k(\alpha) = 1/\pi$, since the Wigner function for the identity is $1/\pi$. The protocol then works by treating all Wigner functions as (proportional to) a genuine probability distribution, which is valid so long as all Wigner functions, $W_\rho, W_k$ are indeed non-negative. In particular, one can draw a phase space sample $\beta$ according to $W_\rho$, and then a measurement outcome $k$ with probability $\pi W_k(\beta)$. 

The above example was for the measurement of a single mode, but the result is easily extended to multi-mode measurements (in effect simply iterating the above each mode). In this case, the computational cost for performing the above procedure is polynomial in the number of modes, provided the distribution $W_\rho$ can be sampled from efficiently (which is typically the case for e.g. pure Gaussian states). 

Let us now try to understand the problem when non-negativity is not guaranteed. First, we make the following proposition:
\begin{proposition}
    There is at most one non-negative Wigner function $W_i\ge 0$ for an orthonormal set $\{\ket{\psi_i}\}_i$, where $W_i$ is the Wigner function for $\ket{\psi_i}$.
    \label{prop:negative-wigner-ortho}
\end{proposition}
\textit{Proof.} Using the second trace condition of Eq.~\eqref{eq:wigner-properties}, we have for $i\neq j$
\begin{equation*}
    0 = |\braket{\psi_i}{\psi_j}|^2 = \pi \int W_i(\alpha)W_j(\alpha)d\alpha.
\end{equation*}
In order for the integral to be exactly 0 requires at least one of $W_i$, $W_j$ to attain negative values (since the Wigner function of a state can not have compact support \cite{WF-support}, we can rule out potential pathological cases such as $W_i, W_j$ having disjoint support on phase space). We can then apply this argument to all pairs $i,j$, which implies either all $W_i$ attain negative values, or there is a single non-negative Wigner function. \hfill \qedsymbol

This implies
\begin{corollary}
    The Wigner function for Fock state $\ket{n}$ attains negative values, for all $n>0$.
\end{corollary}
\textit{Proof.} Follows immediately from the above proposition, noting that $\ket{0}$ has a Gaussian Wigner function. \hfill \qedsymbol

Now let's look at the example of Fock basis measurement on the state\footnote{This is actually a trivial example for a couple of reasons, but it helps illustrate the point. In fact, we show below that one can `cure' the sign problem in this instance, but only because it is a special case, and the result does not (seemingly) generalize.} $\ket{0}$.
Clearly the measurement result `0' should occur with probability 1. Indeed, one has $\pi \int W_0(\alpha)W_n(\alpha)d\alpha = \delta_{n,0}$, where $W_n$ is the Wigner function for $\ket{n}$ (see Eq.~\eqref{eq:fock-wigner}). Now imagine we try to apply the above, first taking a sample according to the genuine distribution $W_0$. Perhaps we pick the value $\beta=1$. Then we can look at the first few values for the Fock states:
\begin{equation*}
    W_0(1) = 0.08615, W_1(1)=0.25847, W_2(1)=0.08615, W_3(1)=-0.20103.
\end{equation*}
Whilst we are still guaranteed $\pi \sum_{n=0}^\infty W_n(1)=1$, there is no obvious (resource efficient) way to map the QPD $\{\pi W_n(1)\}_n$ into the correct distribution (i.e. measurement outcome `0' with probability 1). We can not for example take the absolute values of the terms, ignore the negative contributions etc. The sign problem is an intrinsic issue, related to the fact that on the `local' scale (i.e. individual points in phase space) there is no connection to the `global' properties that contain information about orthogonality, such as $\int W_0 W_1=0$. 
As far as we know, there is no reliable or accurate (i.e. with boundable error) way around this issue, in general, with reasonable resources.

Indeed, `curing' the sign problem is equivalent to finding a unitary $U$ such that $U \rho U^\dag$ as well as $U M_k U^\dag$ all have non-negative Wigner functions, which in almost all practical cases of interest, is infeasible. In fact, by Prop.~\ref{prop:negative-wigner-ortho}, this is not possible to do exactly for an orthonormal basis, as unitarity preserves orthonormality. It is however possible to approximately remove the negativity from e.g. Fock basis measurements, via the `intertwiner' $U = \sum_n \ketbra{\alpha_n}{n}$, where $\braket{\alpha_n}{\alpha_m}\approx \delta_{n,m}$ (so that $U^\dag U \approx \mathbb{I}$, approximately preserving $P(k)$). That is, we map each Fock element to a distinct coherent basis element, such that they remain approximately orthonormal. In general though this will cause the state $U\rho U^\dag$ to be non-Gaussian (unless it also happens to be a Fock basis element, but this is a trivial example). We do see however, this allows one to trade off negativity in the measurements to negativity in the state (approximately).

When the above efficient sampling protocol can not be implemented due to negative values of the Wigner function, one can still attempt to estimate probability outcomes directly, $P_k$, which in turn can allow for sampling from the correct distribution. Though this is typically discussed in the finite dimensional case \cite{quantifying-speedup}, it is acknowledged that the same techniques can be applied in infinite dimensional systems \cite{markov-qpd, lim_approximating_2022}, though of course with slightly different implementation details. We will outline briefly this general approach.

In contrast to the efficient protocol outlined above, one can still sample from the distribution $\{P_k\}_k$ (that is, output measurement `$k$' with probability $P_k$), provided one is able to estimate the $P_k$ to high enough accuracy. In our case, we can re-cast this estimate as a sampling problem itself
\begin{equation}
    P_k = \pi  \int  W_k(\alpha) W_\rho(\alpha)d\alpha = \pi \mathcal{N}_\rho \int  \mathrm{sgn}(W_\rho(\alpha))W_k(\alpha) \frac{|W_\rho(\alpha)|}{\mathcal{N}_\rho}d\alpha,
\end{equation}
where $\mathcal{N}_\rho = \int |W_\rho(\alpha)|d\alpha$ is the Wigner negativity. This implies, if one can sample from the genuine probability distribution $\{ {|W_\rho (\alpha)|}/{\mathcal{N}_\rho } \}_\alpha$, an estimate for $P_k$ can be given:
\begin{equation}
    \tilde{P}_k = \frac{\pi \mathcal{N}_\rho }{N} \sum_{i=1}^N  \mathrm{sgn}(W_\rho(\alpha_i)) W_k(\alpha_i)
\end{equation}
for $N$ samples $\{\alpha_i\}_{i=1}^N$. As shown in Refs.~\cite{markov-qpd, quantifying-speedup}, in order to obtain an estimate, such that with high probability $|P_k - \tilde{P}_k|< \epsilon$, requires a number of samples scaling as $N \propto \mathcal{N}_\rho^2 / \epsilon^2$. For product states, $\rho = \otimes_{i=1}^n \rho_0$, this results in a scaling exponential in $n$\footnote{The Wigner function of a tensor product is a product of the Wigner functions, $W_{A\otimes B}(\alpha, \beta) = W_A(\alpha)W_B(\beta)$.\label{wf-fn}}: $N\propto \mathcal{N}_{\rho_0}^{2n} / \epsilon^2$. For single photon Fock states, $\ket{1}$, one can show (App.~\ref{sect:fock-wigner-neg}), $\mathcal{N}_{\ket{0}} = (4/\sqrt{e}-1)$, resulting in the number of samples scaling approximately as $2^{1.02 n}$.
These results can be improved upon however by (for example) picking a different $s$-parameterized QPD, in effect to minimize the negativity in the representation \cite{lim_approximating_2022}.

With the above method to estimate $P_k$, one can then sample from the distribution $\{P_k\}_k$ in a variety of ways, such as the Markov chain approach or conditional probability sampling as outlined in Sect.~\ref{sect:measurements} (one would need to modify the above construction slightly to work with conditional probabilities, but this is not a problem).
There are however some obvious practical issues related to the huge sampling overhead and errors in the estimation, for example in the case the $P_k$ are exponentially small (as may be the case in Boson sampling), which likely will render such methods impractical, in favor of other simulation techniques as discussed in the main text.

\section{Derivation of Eq.~\eqref{eq:wigner-off-diagonal} \label{sect:derviation-off-diagonal-wigner}}
We wish to compute the Wigner function for off-diagonal coherent state operator $\ketbra{\alpha}{\beta}$. Note, since the Wigner function is linear, in the sense $W_{A+B}= W_A + W_B$ (due to linearity of the trace), this allows us to compute the Wigner function for an arbitrary superposition, $\ket{\psi}=\sum_i c_i\ket{\alpha_i}$, using the density matrix formalism:
\begin{equation*}
    W_{|\psi\rangle} = \sum_{i,j} c_i \bar{c}_j W_{\ketbra{\alpha_i}{\alpha_j}}.
\end{equation*}
Also note that since the Wigner function of a tensor product is a product of Wigner functions, $W_{A\otimes B}(\alpha, \beta) = W_A(\alpha) W_B(\beta)$, this allows one to compute the Wigner function for multi-mode coherent state superpositions, as of interest in the present work, though here we'll focus on the single mode case.

Using Eq.~\eqref{eq:wigner}, we first need to evaluate the characteristic function $\mathrm{Tr}[\hat{D}(\gamma)\ketbra{\alpha}{\beta}]$. We will require the identity $\hat{D}(\gamma)\ket{\alpha} = e^{i\mathrm{Im}(\bar{\alpha} {\gamma})}\ket{\alpha+\gamma}$, as well as\footnote{$\mathrm{Tr}[\ketbra{\alpha}{\beta}] = \sum_{n\ge 0} \langle n|\alpha \rangle \langle \beta |n\rangle =  \sum_{n\ge 0} \langle \beta |n\rangle \langle n|\alpha \rangle =\braket{\beta}{\alpha}$, by resolution of the identity in the Fock basis.} $\mathrm{Tr}[\ketbra{\alpha}{\beta}]=\braket{\beta}{\alpha}$. Then
\begin{equation}
    \mathrm{Tr}[\hat{D}(\gamma)\ketbra{\alpha}{\beta}] = e^{i\mathrm{Im}(\alpha \bar{\gamma})}\braket{\beta}{\alpha + \gamma} = e^{i\mathrm{Im}(\bar{\alpha} {\gamma})} e^{-i\mathrm{Im}(\beta(\bar{\alpha} + \bar{\gamma}))}e^{-|\alpha + \gamma - \beta|^2/2}
\end{equation}
where the last step uses Eq.~\eqref{eq:coherent-overlap}. The goal is to now compute
\begin{equation}
    W_{\ketbra{\alpha}{\beta}}(\kappa) =\frac{1}{\pi^2} \int d\gamma e^{\bar{\gamma}\kappa - \gamma \bar{\kappa}}e^{i\mathrm{Im}(\bar{\alpha} {\gamma})} e^{-i\mathrm{Im}(\beta(\bar{\alpha} + \bar{\gamma}))}e^{-|\alpha + \gamma - \beta|^2/2},
\end{equation}
where $d\gamma = d\gamma_r d\gamma_i$, where $\gamma = \gamma_r + i \gamma_i$ (we'll use the same notation for the other parameters). The integration limits are both $(-\infty, \infty)$.
Since this is the product of two integrals, we can collect the $\gamma_r$ and $\gamma_i$ terms separately. The separate integrands are:
\begin{equation*}
    e^{2i\gamma_r \kappa_i}e^{-i\gamma_r \beta_i}e^{-(\gamma_r + \alpha_r - \beta_r)^2/2};\; e^{-2i\gamma_i \kappa_r}e^{i\gamma_i \beta_r}e^{-(\gamma_i + \alpha_i - \beta_i)^2/2}.
\end{equation*}
This also leaves a constant $e^{i(\beta_r\alpha_i - \beta_i \alpha_r)}$. Multiplying out the square terms gives us a product of two Gaussian integrals:
\begin{equation*}
    W_{\ketbra{\alpha}{\beta}}(\kappa) =\frac{1}{\pi^2} e^{i(\beta_r\alpha_i - \beta_i \alpha_r)}e^{-\frac{1}{2}|\alpha-\beta|^2} \int d\gamma_r e^{-\frac{1}{2}\gamma_r^2 + \gamma_r(\beta_r-\alpha_r + 2i\kappa_i - \beta_i)}\int d\gamma_i e^{-\frac{1}{2}\gamma_i^2 + \gamma_i(\beta_i-\alpha_i - 2i\kappa_r + \beta_r)}
\end{equation*}
which, using the result, for $a,b\in \mathbb{C}$, so long as $\mathrm{Re}(a)>0$, $\int dx e^{-ax^2 + bx}=\sqrt{\pi/a}e^{b^2/4a}$ we get, with some light rearranging:
\begin{equation}
    W_{\ketbra{\alpha}{\beta}}(\kappa) = \frac{2}{\pi}e^{i\phi_{\alpha,\beta}(\kappa)} e^{-2|\kappa - \frac{1}{2}(\alpha + \beta)|^2} = e^{i\phi_{\alpha, \beta}(\kappa)}W_{\ket{\frac{1}{2}(\alpha+ \beta)}}(\kappa),
\end{equation}
where 
\begin{equation*}
    \phi_{\alpha, \beta}(\kappa) = 2\kappa_i(\beta_r - \alpha_r) + 2\kappa_r(\alpha_i - \beta_i) + \alpha_r \beta_i - \alpha_i \beta_r.
\end{equation*}
That is, the Wigner function for $\ketbra{\alpha}{\beta}$ is the Wigner function for the coherent state $\ket{\frac{1}{2}(\alpha+ \beta)}$, multiplied by a phase factor.

Some comments are in order:
\begin{itemize}
    \item Setting $\alpha=\beta$, we recover the Wigner function for $\ket{\alpha}$, as required ($\phi_{\alpha,\alpha}=0$)
    \item For a cat state, with $\beta = -\alpha$, the phase factor simplifies to $\phi = 4(\kappa_r\alpha_i - \kappa_i\alpha_r)$
    \item Since $\int W_A(\kappa)d\kappa = \mathrm{Tr}[A]$, we have $\int W_{\ketbra{\alpha}{\beta}}(\kappa)d\kappa = \braket{\beta}{\alpha} = \int e^{i\phi_{\alpha, \beta}(\kappa)}W_{\ket{\frac{1}{2}(\alpha+\beta)}}(\kappa)d\kappa$
    \item Notice $\phi_{\alpha, \beta} = -\phi_{\beta,\alpha}$, which implies $\overline{W}_{\ketbra{\alpha}{\beta}}(\kappa) = e^{-i\phi_{\alpha, \beta}(\kappa)}W_{\ket{\frac{1}{2}(\alpha+\beta)}}(\kappa) = W_{\ketbra{\beta}{\alpha}}(\kappa)$
\end{itemize}
This last point in particular implies, for a state $\ket{\psi}=\sum_i c_i \ket{\alpha_i}$, the Wigner function is (dropping the explicit $\kappa$ dependence for simplicity)
\begin{equation}
    W_{|\psi\rangle} = \sum_{i,j} c_i \bar{c}_j W_{\ketbra{\alpha_i}{\alpha_j}} = \sum_i |c_i|^2 W_{\ket{\alpha_i}} + 2\sum_{j>i} W_{\ket{\frac{1}{2}(\alpha_i+\alpha_j)}}\mathrm{Re}\left[c_i \bar{c}_j e^{i\phi_{\alpha_i, \alpha_j}} \right],
\end{equation}
and thus all negativity in the Wigner function, if any, comes from the terms $\mathrm{Re}\left[c_i \bar{c}_j e^{i\phi_{\alpha_i, \alpha_j}} \right]$, related to the coherences. If the amplitudes are all real, then this simplifies to
\begin{equation}
    W_{|\psi\rangle} =  \sum_i c_i^2 W_{\ket{\alpha_i}} + 2\sum_{j>i}c_i c_j W_{\ket{\frac{1}{2}(\alpha_i+\alpha_j)}}\cos \phi_{\alpha_i, \alpha_j}.
\end{equation}

\section{Squeezed state decomposition to coherent states \label{sect:squeeze-approx}}
Since a squeezed vacuum state Eq.~\eqref{eq:squeezed_state} is a superposition over all even Fock states, it suggests to approximate a squeezed state by a superposition of even cat states,
\begin{equation*}
   \ket{\zeta} \approx \sum_{k=0}^N c_k (\ket{\alpha_k} + \ket{-\alpha_k}) =: \ket{\tilde{\zeta}},
\end{equation*}
where $\zeta = re^{i\phi}$. Here $\ket{\zeta}$ should not be confused with a coherent state, this is simply a short coming of the notation. The $\ket{\alpha_k}$ on the right hand side are however coherent states.
Such a decomposition also has the property that all odd Fock states have identically 0 amplitude. We posit this form, and follow a similar strategy as in the proof of Th.~\ref{th:general-fock}.

In particular, if we desire to approximate a squeezed state up to $2N$ photons, we can pick $\alpha_k = e^{\pi i k / (N+1)}\sqrt{-\frac{1}{2a}e^{i\phi}\tanh r}$, where for now $a$ is a free parameter. With this, we get the following relation for the amplitude on $\ket{2n}$,
\begin{equation*}
    \sum_{k=0}^N c_k e^{2\pi i n k / (N+1)} = a^n\sqrt{\frac{e^{\frac{1}{2a}\tanh r}}{4{\cosh r}}} \frac{\sqrt{(2n)!}}{n!}=:f_n,
\end{equation*}
valid for $n=0, \dots, N$.
Using the same Fourier analysis as in Th.~\ref{th:general-fock}, one can show the amplitudes must satisfy
\begin{equation}
    c_k = \frac{1}{N+1}\sum_{n=0}^N f_n e^{-2\pi i n k/(N+1)}.
\end{equation}

Such a construction gives the correct amplitudes, exactly, up to Fock state $\ket{2N+1}$.
We can in fact do slightly better than this, as we have still got the parameter $a$ unspecified. This can be utilized in two ways. First, by taking $a$ large enough, we can guarantee the amplitudes in the approximate state are bounded for $n>2N$. Another approach is to use $a$ to exactly reproduce also the next amplitude, for $\ket{2N+2}$. We can do this since, the amplitude on $\ket{2N+2}$ is
\begin{equation}
    \langle 2N+2|\tilde{\zeta}\rangle = \frac{1}{\sqrt{(2N+2)!}}\left(-\frac{1}{2a}e^{i\phi}\tanh r\right)^{N+1}  \frac{1}{\sqrt{\cosh r}},
\end{equation}
which can be re-arranged to solve for $a$ so that the right hand side equals the desired amplitude.
With this, using a superposition of $2N+2$ coherent states ($N+1$ cat states), the approximate (unnormalized) state $\ket{\tilde{\zeta}}$ has the correct amplitudes up to level $\ket{2N+3}$.

Upon appropriate normalization of $\ket{\tilde{\zeta}}$, so long as enough terms are taken, an arbitrary accuracy can be achieved. In practice, as shown in Fig.~\ref{fig:squeeze_approx}, for small enough $r$, very few terms are needed, but large $r$ could require an impractical number of coherent states.

\section{A different proof of Cor.~\ref{cor:creation_cost} using postselection \label{sect:alternate-proof-adag}}

\begin{proposition}
    The resourceful operation $\hat{a}^\dag$ can be implemented on a rank $k$ single-mode state $\ket{\psi}$ to arbitrary accuracy, resulting in  approximate coherent rank $2k$. This is achieved via
    \begin{equation*}
        \hat{a}^\dag \ket{\psi} = \frac{1}{\epsilon}(\mathbb{I}\otimes \bra{0})\hat{B}(2\epsilon, 0)\ket{\psi,1} + O(\epsilon^2),
    \end{equation*}
    where $\hat{B}(\theta,\phi)$ is the beamsplitter unitary, Eq.~\eqref{eq:beamsplitter}.
\end{proposition}
\textit{Proof.} Consider the beamsplitter unitary of Sect.~\ref{sect:free-ops}, $\hat{B}(2\epsilon, 0) = e^{\epsilon (\hat{a}^\dag \hat{b} - \hat{a} \hat{b}^\dag)}$. Notice, to first order in $\epsilon$, applying $\frac{1}{\epsilon}\hat{B}(2\epsilon,0)$ on $\ket{\psi,1}$ yields
\begin{equation}
    \frac{1}{\epsilon} B(2\epsilon, 0) \ket{\psi, 1} = \frac{1}{\epsilon} \ket{\psi, 1} +  \hat{a}^\dag \ket{\psi} \ket{0} + \sqrt{2} \hat{a}\ket{\psi} \ket{2} + O(\epsilon).
    \label{eq:creation_beamsplitter}
\end{equation}
Since $\ket{1}$ can be written to arbitrary accuracy using $2$ coherent states (Cor.~\ref{cor:fock}), $\ket{\psi,1}$ has approximate coherent state rank $2k$. Then, as the beamsplitter is a free operation (Sect.~\ref{sect:free-ops}), Eq.~\eqref{eq:creation_beamsplitter} also has rank $2k$.
The final step is to perform a projection onto $\ket{0}$ on the auxiliary mode, which is a free operation (e.g. see Sect.~\ref{sect:measurements}).

The error in this approximation arises from higher order terms ($O(\epsilon^2)$ or greater) where the auxiliary mode is also in the $\ket{0}$ state (e.g. $(\hat{a}^\dag \hat{b})^2 \hat{a}\hat{b}^\dag$), but this can be made arbitrarily small by taking $\epsilon \rightarrow 0$.

We have thus shown to error $O(\epsilon^2)$, we can implement $\hat{a}^\dag\ket{\psi}$, yielding an approximate coherent rank $2k$ state.
\hfill \qedsymbol

The above can be trivially applied to $\hat{a}^\dag$ acting on any mode of a multi-mode state, with the same complexity. A similar trick can also be used to compute $(\hat{a}^\dag)^n\ket{\psi}$, by applying the beamsplitter (normalized by $\sqrt{n!}/\epsilon^n$) on initial state $\ket{\psi, n}$, and projecting the auxiliary mode on $\ket{0}$ (in this case, the rank increases by a factor of $n+1$).

\section{Wigner negativity for Fock states $\ket{1}, \ket{2}$ \label{sect:fock-wigner-neg}}

The Wigner function for a Fock state $\ket{n}$ is \cite{gerry_knight_2004}
\begin{equation}
    W_n(\alpha)= (-1)^n\frac{2}{\pi}e^{-2|\alpha|^2}L_n(4|\alpha|^2)
    \label{eq:fock-wigner}
\end{equation}
where $L_n$ is the $n$'th Laguerre polynomial. 
For $n=0,1,2$ we have
\begin{equation}
    L_0(x)=1;\;L_1(x) = 1-x;\; L_2(x) = \frac{1}{2}(x^2 - 4x + 2).
\end{equation}

Here we analytically compute the integral of the absolute value of $W_{1,2}$, to use in comparison with our metric of coherent state rank monotone, as in Sect.~\ref{sect:resource_th}. The first step is to convert to polar coordinates, $\alpha = r e^{i\theta}$. Then
\begin{equation*}
    \int_{\mathbb{R}^2} |W_n(\alpha)|d\alpha = 4 \int_0^\infty e^{-2r^2}|L_n(4r^2)|r dr,
\end{equation*}
where $d\alpha = d\mathrm{Re}(\alpha)d\mathrm{Im}(\alpha)$.

For $n=1$ in particular, this gives (noting the integrand has a root at $1/2$)
\begin{equation}
    \int_{\mathbb{R}^2} |W_1(\alpha)|d\alpha = 4 \int_0^{1/2} e^{-2r^2}(r - 4r^3)dr + 4\int_{1/2}^\infty e^{-2r^2}(4r^3-r)dr.
\end{equation}
One can readily check that $\int_a^b e^{-2r^2}(1-4r^2)rdr = \frac{1}{4}e^{-2r^2}(4r^2+1)|_a^b$, to show
\begin{equation}
    \int_{\mathbb{R}^2} |W_1(\alpha)|d\alpha = 4e^{-1/2} - 1 \approx 1.42612264.
\end{equation}
That is, the logarithm of the Wigner negativity is: $\mathcal{W}(1)=\log_2 (4e^{-1/2}-1) \approx 0.51209805$. We can compare this to $\mathcal{R}(1)=\log_2 2 = 1$.

For $n=2$, the integrand becomes $e^{-2r^2}(8r^4 - 8r^2 + 1)r$, which has roots (for $r>0$) at $r_\pm = \frac{1}{\sqrt{2}}\sqrt{1 \pm 1/\sqrt{2}}$. This gives integration limits $\int_0^{r_-} - \int_{r_-}^{r_+} + \int_{r_+}^\infty$, taking into account the modulus. Using that $\int_a^b e^{-2r^2}(8r^4 - 8r^2 + 1)r = -\frac{1}{4}e^{-2r^2}(8r^4 + 1)|_a^b$, one can easily check that
\begin{equation}
    \int_{\mathbb{R}^2}|W_2(\alpha)|d\alpha = 8e^{-1}\left(\sqrt{2} \cosh \frac{1}{\sqrt{2}} - 2 \sinh \frac{1}{\sqrt{2}}\right) + 1 \approx 1.72898926.
\end{equation}
Likewise, $\mathcal{W}(2) \approx 0.78992891$, and $\mathcal{R}(2)=\log 3 \approx 1.58496250$.

The ratios $\mathcal{R}(1)/\mathcal{W}(1)\approx 1.95275104$ and $\mathcal{R}(2)/\mathcal{W}(2)\approx 2.00646221$ imply there is not a simple condition relating $\mathcal{W}$ to $\mathcal{R}$, although as shown in Fig.~\ref{fig:fock-resource}, $\mathcal{W}(n)\approx 0.48 \mathcal{R}(n)$ holds for the first few $n$. The growth of $\mathcal{W}_n$ can most likely be attributed to the fact that $W_n$ has $n$ roots (i.e. the number of negative regions grows linearly in $n$).

\section{Free unitaries cannot consume quantum resources}
\label{sec:free-unitaries}
Let \(\mathcal{U}_\mathcal{F}\) denote the set of free unitaries. For \(U \in \mathcal{U}_{\mathcal{F}}\) and all \(\rho \in \mathcal{F}\), we have, by definition
 \begin{align}
 U \rho U^{\dagger} = \sigma \in \mathcal{F}.
 \end{align}
 Inverting the unitary implies that,
 \begin{align}
 U^{\dagger} \sigma U = \rho \in \mathcal{F} ~~\forall \sigma \in \mathcal{F}.
 \end{align}
 Therefore, \(U^{\dagger} \in \mathcal{U}_{\mathcal{F}}\) as well.

 Now, consider the action of the free unitary on a resourceful state, \(\omega \not \in \mathcal{F}\). We want to see if \(U \in \mathcal{U}_{\mathcal{F}}\) can \textit{consume} the resource in this state. I.e.,
 \begin{align}
 \mathcal{R}(U \omega U^{\dagger}) < \mathcal{R}(\omega)?
 \end{align}
 (Note the strict inequality above). Define \(\kappa := U \omega U^{\dagger}\) then, using monotonicity we know, \(\mathcal{R}(\kappa) \leq \mathcal{R}(\omega)\). However, since \(U^{\dagger}\) is a free operation as well, we have \(\mathcal{R}(U^{\dagger} \kappa U) \leq \mathcal{R}(\kappa)\) (using monotonicity of \(U^{\dagger}\)). Since \(U^{\dagger} \kappa U = \omega\), this is equivalent to, \(\mathcal{R}(\omega) \leq \mathcal{R}(\kappa)\). Combining the two inequalities, we must have, \(\mathcal{R}(U \omega U^{\dagger}) = \mathcal{R}(\omega)\) for \(\omega \not \in \mathcal{F}\). That is, free unitaries cannot be used to \textit{consume} any resource (e.g., in quantum teleportation, measurements are necessary to consume entanglement). We need \textit{nonunitary} operations, e.g., measurement, partial trace, etc., to consume resources. That is, manipulation of quantum resources requires quantum \textit{channels}. 
 
As a simple example from pure, bipartite entanglement theory, one can quantify the amount of resource in a state via the von Neumann entropy of the reduced state. And local unitaries (that are free operations) cannot change the von Neumann entropy, while local measurements (which are also free) can, e.g., in quantum teleportation. In fact, this is the reason why the set of free operations \(\mathcal{O}\) is usually assumed to have a \textit{semigroup} structure and \textit{not} a group structure, since state transformation is trivialized under the latter.
\end{document}